\documentclass[12pt]{article}
\usepackage{multirow}
\usepackage{graphicx}
\usepackage{graphics}
\usepackage{epsfig}
\usepackage[below]{placeins}
\usepackage{subfigure}
\usepackage{float}
\usepackage{rotating}
\usepackage{amsmath}
\usepackage{longtable}
\usepackage{booktabs}
\usepackage{subfigure}

\begin{document}

\begin{center}

{\bf \Large A Fully Automated Latent Fingerprint Matcher with Embedded Self-learning Segmentation Module}

\bigskip

{\bf Jinwei Xu, Jiankun Hu, Xiuping Jia}\\

School of Engineering and Information Technology\\

The University of New South Wales\\

Canberra, ACT 2600, Australia\\

E-mails: jinwei.xu@student.adfa.edu.au; j.hu@adfa.edu.au; x.jia@adfa.edu.au

\end{center}

\begin{center}
{\bf Abstract}
\end{center}

Latent fingerprint has the practical value to identify the suspects who have unintentionally left a trace of fingerprint in the crime scenes. However, designing a fully automated latent fingerprint matcher is a very challenging task as it needs to address many challenging issues including the separation of overlapping structured patterns over the partial and poor quality latent fingerprint image, and finding a match against a large background database that would have different resolutions. Currently there is no fully automated latent fingerprint matcher available to the public and most literature reports have utilized a specialized latent fingerprint matcher COTS3 which is not accessible to the public. This will make it infeasible to assess and compare the relevant research work which is vital for this research community. In this study, we target to develop a fully automated latent matcher for adaptive detection of the region of interest and robust matching of latent prints. Unlike the manually conducted matching procedure, the proposed latent matcher can run like a sealed black box without any  manual intervention. This matcher consists of the following two modules: (i) the dictionary learning-based region of interest (ROI) segmentation scheme; and (ii) the genetic algorithm-based minutiae set matching unit. Experimental results on NIST SD27 latent fingerprint database demonstrates that the proposed matcher outperforms the currently public state-of-art latent fingerprint matcher.

\bigskip


\section{Introduction}

%------------------- example for sub-figures -------------------
%\begin{figure*}
%    \centering
%    \subfigure[XXX - First caption]
%    {
%        \includegraphics[scale=0.2]{CMC_on_Different_Methods.pdf}
%        \label{fig:first_sub}
%    }
%    \subfigure[YYY - Second caption]
%    {
%        \includegraphics[scale=0.2]{CMC_on_Different_Methods.pdf}
%        \label{fig:second_sub}
%    }
%    \subfigure[ZZZ - Third caption]
%    {
%        \includegraphics[scale=0.2]{CMC_on_Different_Methods.pdf}
%        \label{fig:third_sub}
%    }
%    \caption{Different $\overline{CMC}$ obtained by the three scenarios}
%    \label{fig:sample_subfigures}
%\end{figure*}
%
%% reference the figure
%Figure \ref{fig:second_sub}
%------------------- example for sub-figures -------------------

The pioneering study for fingerprint identification with its application in distinguishing criminals could be traced back to \cite{Faulds1880}. The fingerprints inadvertently touched by a person in crime scenes is applicable to identify the criminals or to exclude the suspects. Fingerprints collected from crime scenes are compared to the fingerprints collected from suspects so that the fingerprints belonging to criminals could be identified \cite{Maltoni09}. As a consequence, the Automated Fingerprint Identification Systems (AFIS) is established and developed to satisfy such urgent need \cite{Komarinski05}. One important function of the AFIS system is to identify suspects against a large fingerprint database from an unknown fingerprint. There are two basic ways of searching. One approach is fingerprint indexing including classification, where the query fingerprint is mapped into a cluster with similar characteristics and such cluster will become the candidates for further inspection. The second approach is to perform matching on a one-on-one basis against the whole database. In principal, the first approach is most efficient. However, even though some progress has been made on the partial fingerprint indexing \cite{Wang11}, few literatures have been found on latent fingerprint indexing. Existing latent fingerprint identification work is virtually on a one-on-one basis which is also our focus in this paper. Therefore in the remaining of the paper, our AFIS discussion is restricted to the category of one-on-one matching unless stated otherwise. AFIS is widely used to identify three main types of fingerprints: the rolled, the plain and the latent. The rolled fingerprint is a print which is obtained by rolling the finger from one side of nail to the other side of nail (namely, nail-to-nail) on a card or inside a platen scanner (shown in Figure \ref{show_different_prints:rolled}). The plain is a print collected by pressing the finger down on a card or place the finger flat on a scanner (shown in Figure \ref{show_different_prints:plain}). The latent ones are acquired from crime scenes where the prints are not intentionally touched by the suspects or criminals (shown in Figure \ref{show_different_prints:latent}). For the rolled and plain fingerprints, both are acquired in a controlled mode. That is, they are typically in good quality and are rich of reliable detailed features (e.g. minutiae). Consequently AFIS is able to handle the rolled and plain identification cases in full-automatic mode. In contrast, the fingerprint in latent images are usually small-sized, overlapped with other image components and blurred due to the following possible causes:

%------------------- show rolled / plain / latent -------------------
\begin{figure}
    \centering
    \subfigure[]
    {
        \includegraphics[height=2in]{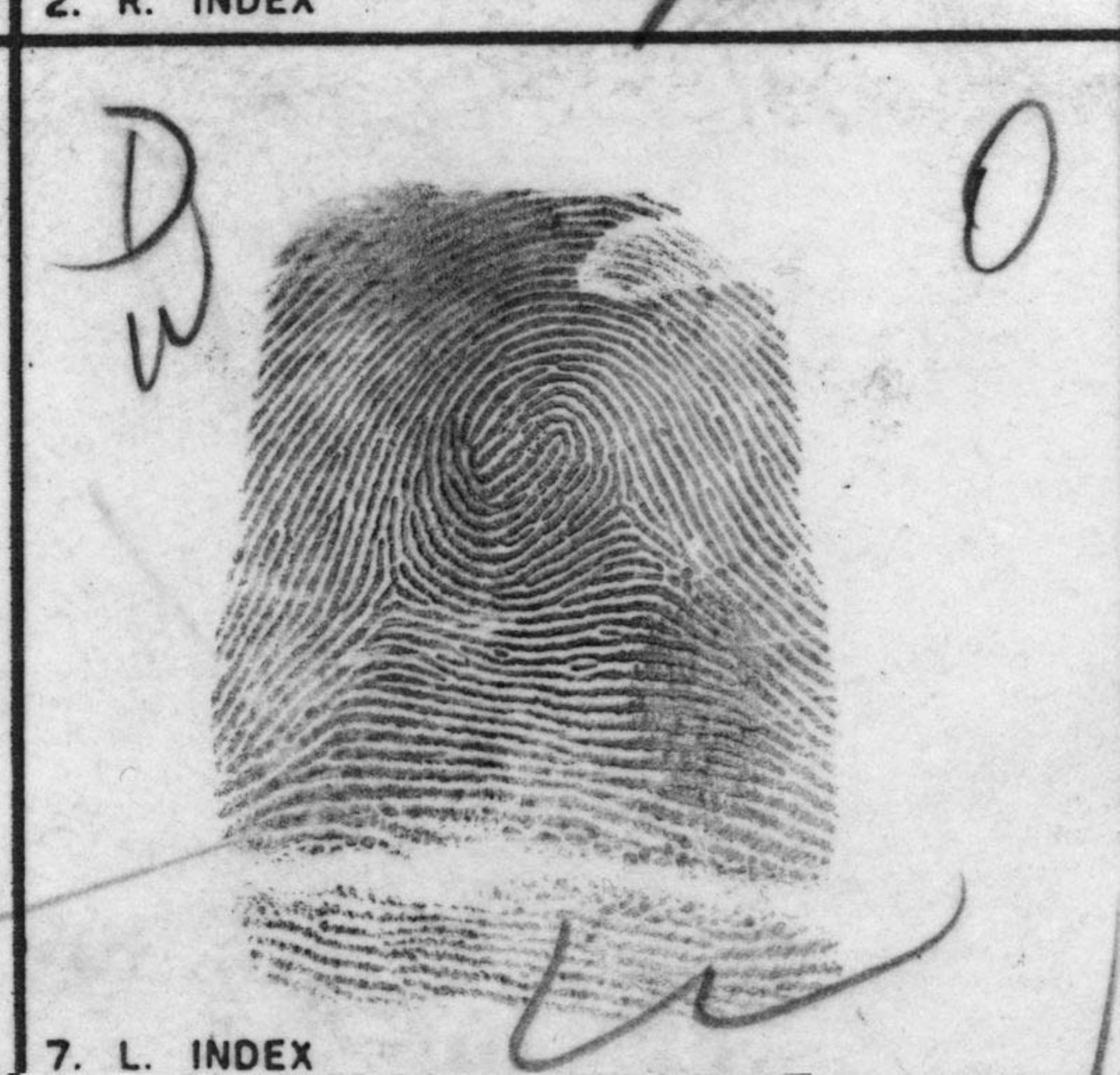}
        \label{show_different_prints:rolled}
    }
    \subfigure[]
    {
        \includegraphics[height=2in]{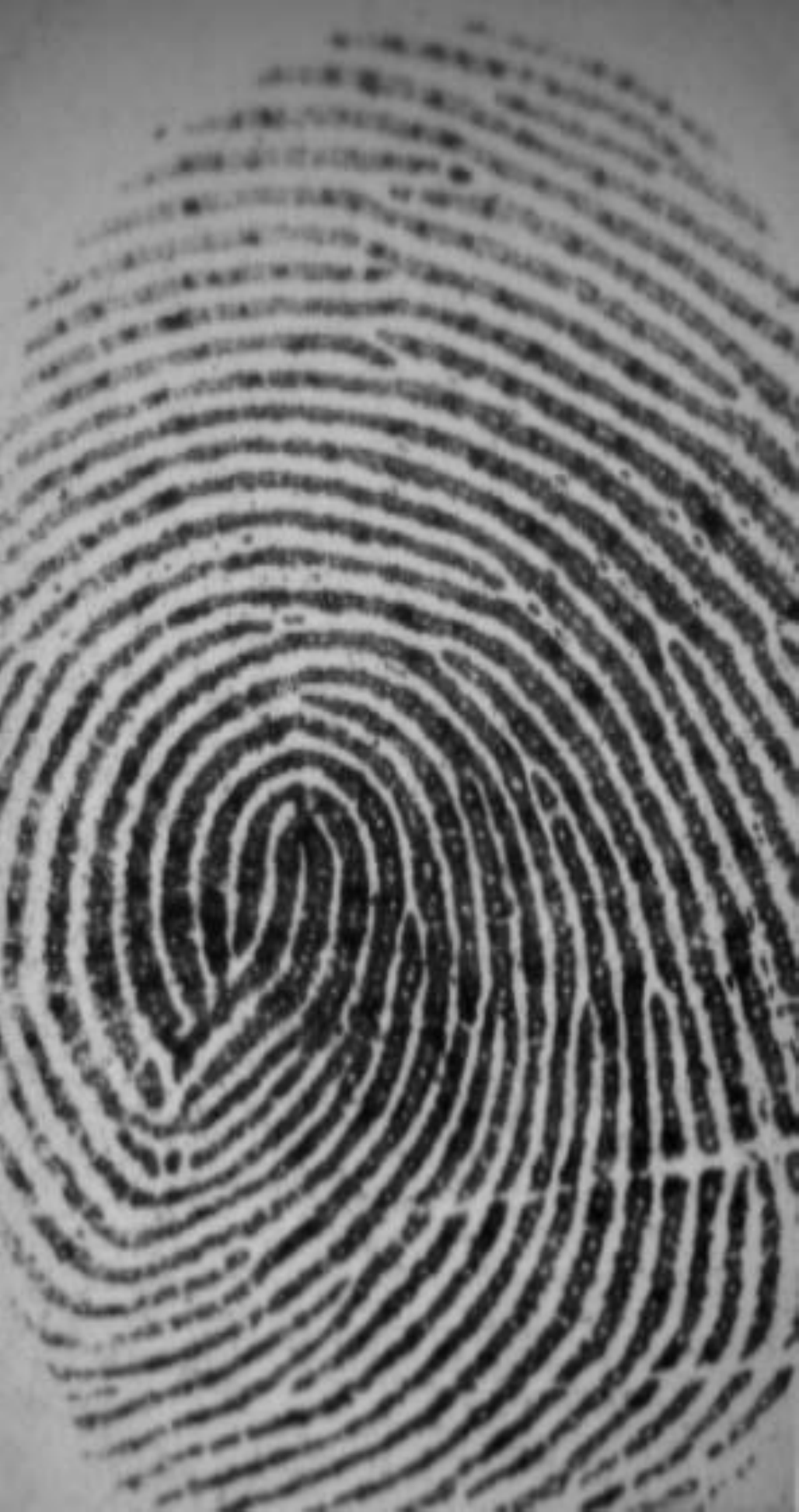}
        \label{show_different_prints:plain}
    }
    \subfigure[]
    {
        \includegraphics[height=2in]{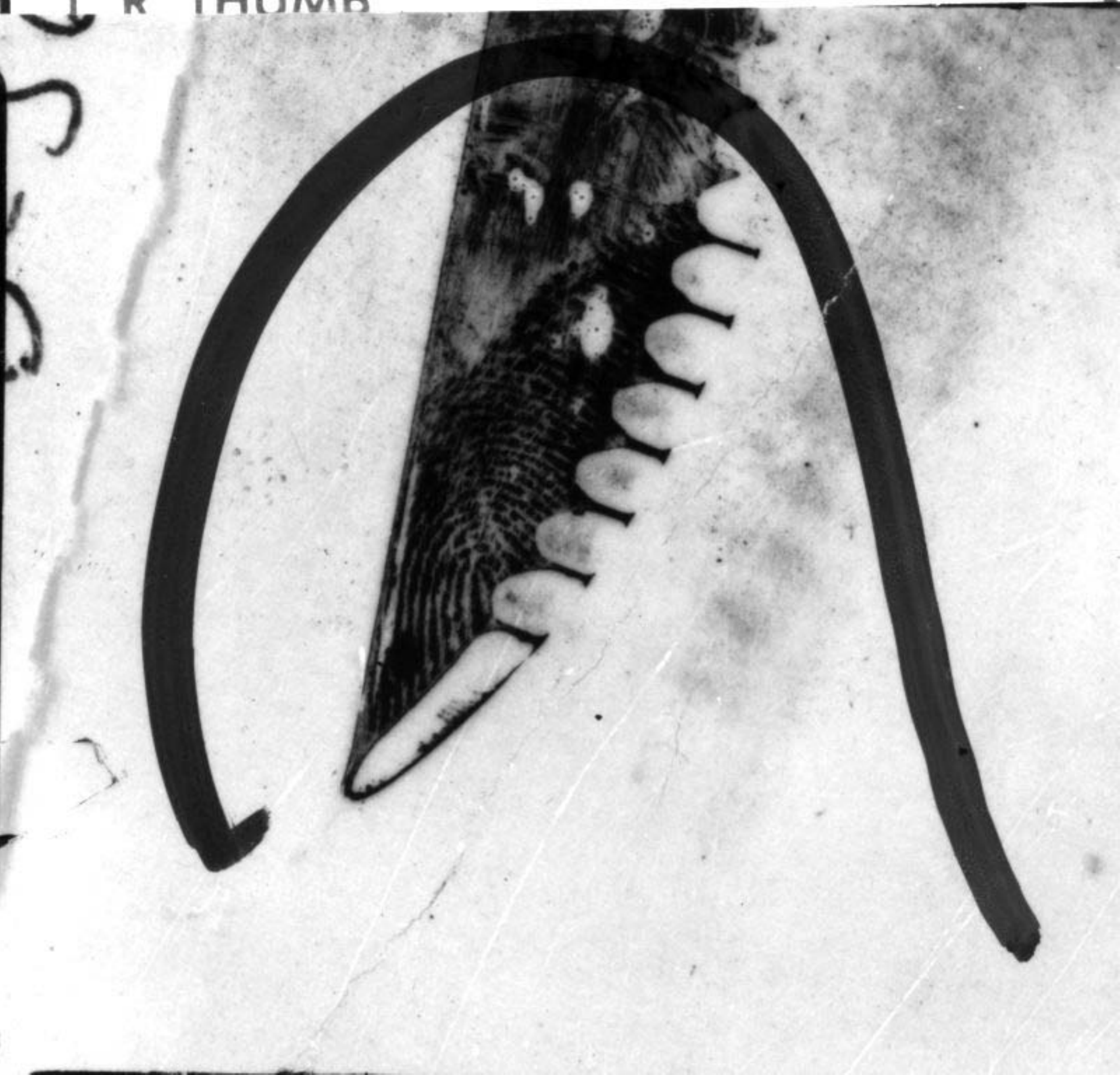}
        \label{show_different_prints:latent}
    }
    \caption{Three types of fingerprints: (a) the rolled print; (b) the plain print; and (c) the latent print.}
    \label{show_different_prints}
\end{figure}
%------------------- show rolled / plain / latent -------------------

%------------------- small / smudge / blur -------------------
\begin{figure}
    \centering
    \subfigure[]
    {
        \includegraphics[height=1.7in]{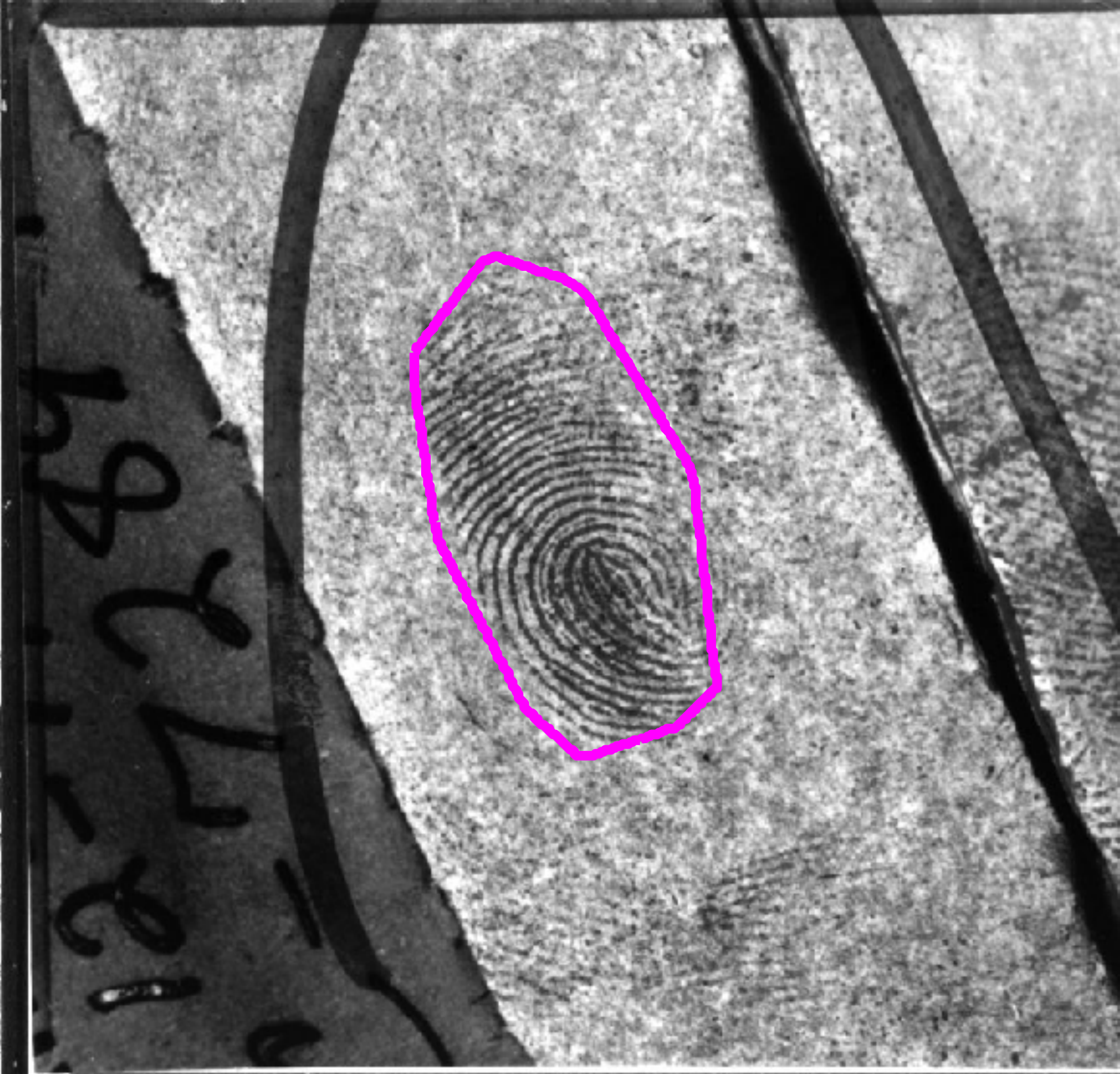}
        \label{show_different_distortion:small_area}
    }
    \subfigure[]
    {
        \includegraphics[height=1.7in]{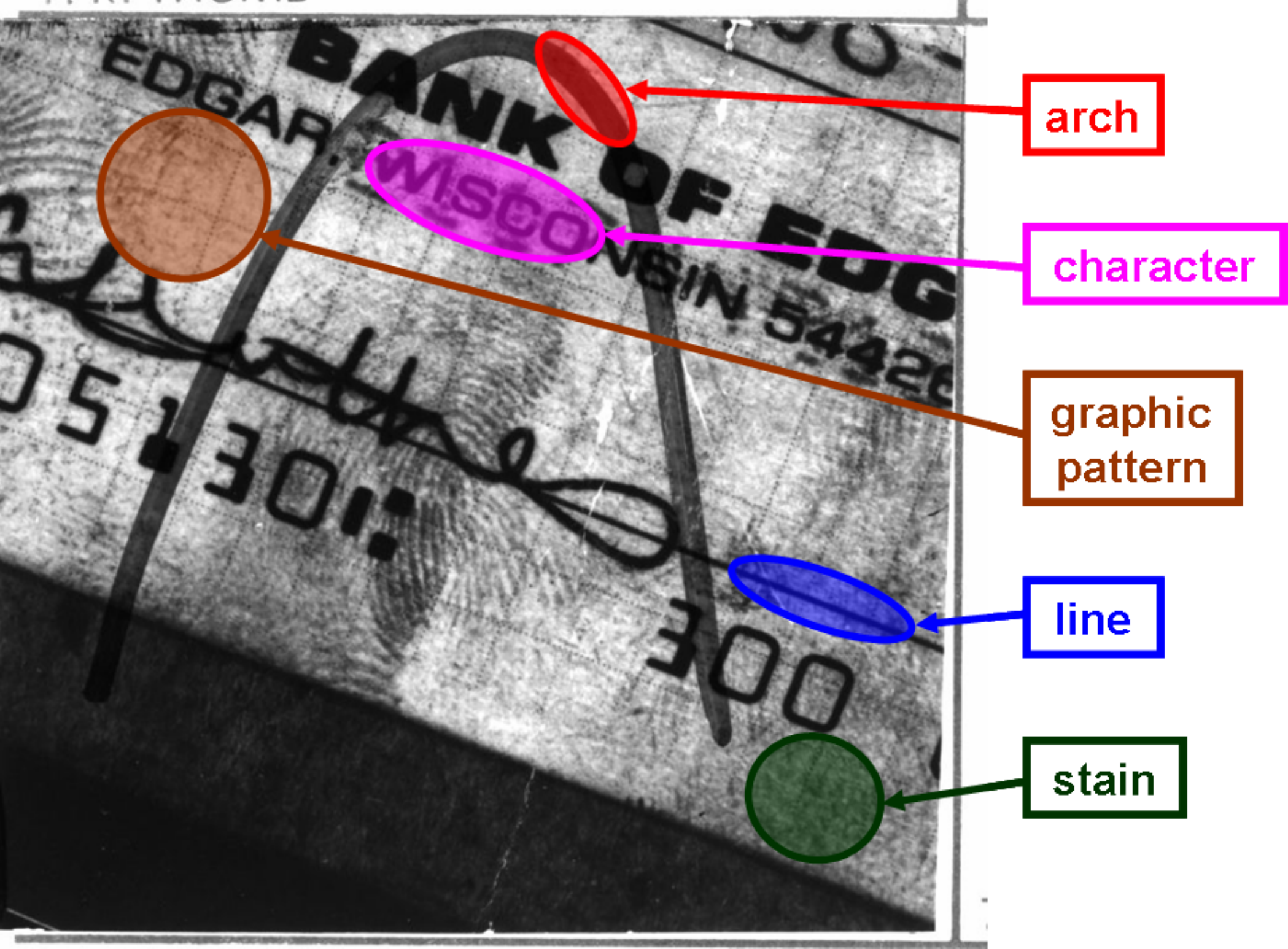}
        \label{show_different_distortion:smudge}
    }
    \subfigure[]
    {
        \includegraphics[height=1.7in]{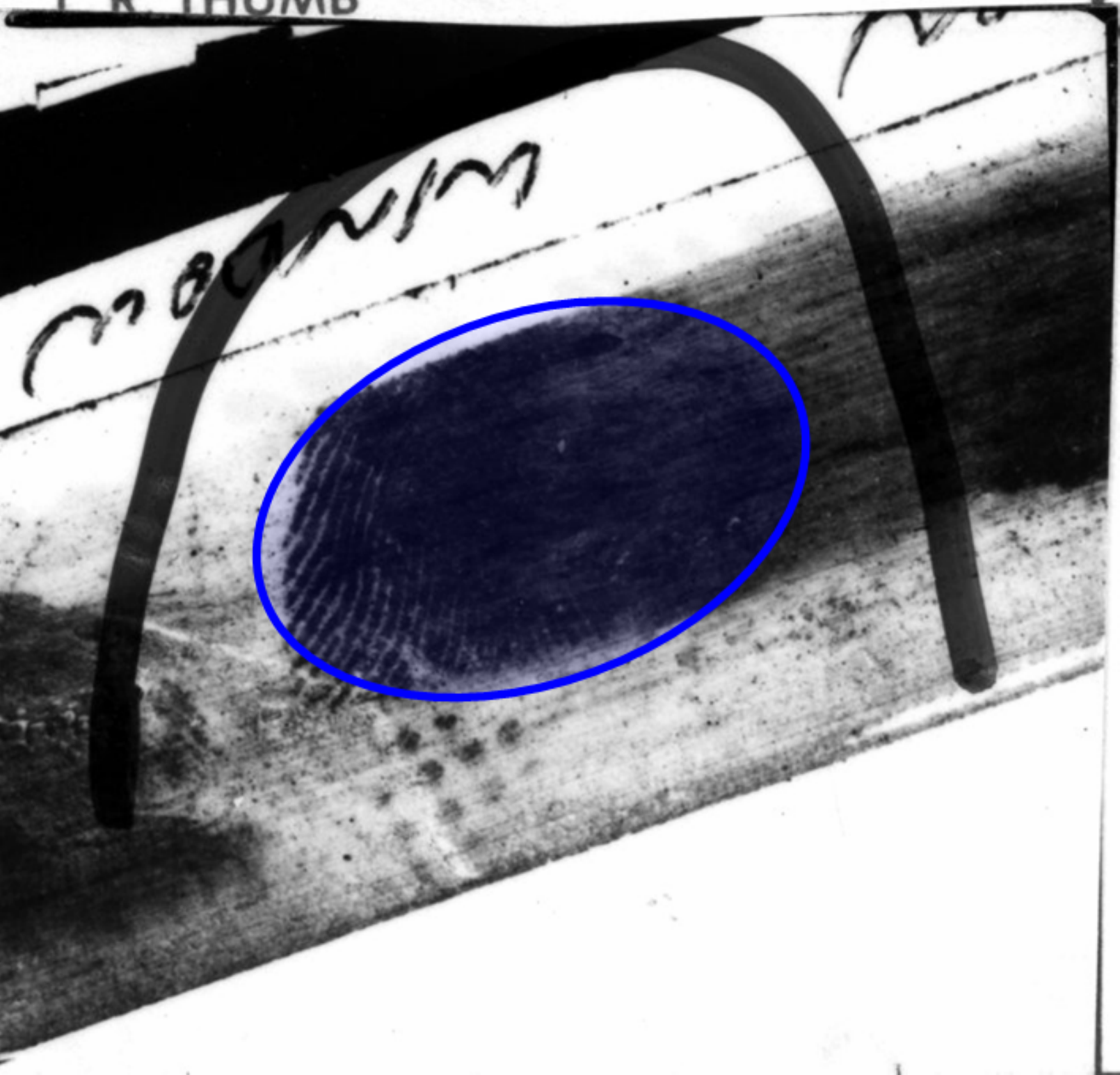}
        \label{show_different_distortion:blur}
    }
    \caption{The challenges involved in latent print images: (a) small area in ROI; (b) overlapping with the various structured noise; and (c) blur.}
    \label{show_different_distortion}
\end{figure}
%------------------- small / smudge / blur -------------------

\begin{itemize}

  \item Small area: the most fingerprints collected from crime scenes are not complete but partial (shown in Figure \ref{show_different_distortion:small_area});

  \item Overlapping with other structured components: the fingerprints usually overlap with other structured noise such as arch, line, character, stain, and graphic pattern (shown in Figure \ref{show_different_distortion:smudge});

  \item Blur: the most fingerprints acquired from crime scenes have large distortion due to the pressure variations when the fingers touching or pressing down the object surface (shown in Figure \ref{show_different_distortion:blur}).

\end{itemize}

All above adverse effects are not solely encountered but concurrently confronted, therefore latent fingerprint images are generally in poor quality. Considering that AFIS primarily depends on the sufficient and reliable features, the latent fingerprint identification based on automatically extracted features is inaccurate. That is, the poor image quality imposes the difficulties on the automated feature extraction so that the extracted features are limited and the most of them are not reliable. Consequently the fully-automated matching conducted by AFIS based on the limited and unreliable features would lead to the inaccurate result. In order to ensure the reliability and accuracy for latent fingerprint identification, the good-quality ROI as well as the reliable features are often manually marked instead of automatically extracted. With the human intervention involved, for latent fingerprint identification, the semi-automatic mode rather than full-automatic mode is adopted.

The semi-automatic latent fingerprint identification procedure consists of the following four stages: (i) the ROI in latent images are manually labeled; (ii) based on the labeled ROI, the features such as minutiae, singularity, ridge quality map, orientation field, ridge wavelength map, and skeleton are manually extracted \cite{Jain08}; (iii) the marked features are uploaded to a latent fingerprint matcher, then are automatically matched against the features derived from the rolled / plain fingerprints in background database; and (vi) according to the matching scores, the candidate rolled / plain prints are retrieved, and the candidates are visually verified by latent examiners. After visual verification, the most possible archived fingerprint in background database might be found. If not found, a new round routine with more cautious ROI labeling, feature extraction, and visual verification would be conducted for obtaining the most possible candidate. For semi-automatic mode in latent fingerprint identification, the resultant accuracy is satisfactory \cite{Jain11}.

In order to reduce the cost of expert work, a fully automated latent fingerprint matching and identification is needed \cite{Cao14}. This is also desirable for the advancement of the technology. \cite{Paulino11} and \cite{Paulino13} propose a latent matching algorithm without involving many manually marked features and it only used the minutiae information provided by latent experts. Compared with \cite{Jain08} and \cite{Jain11}, the most elaborately marked features (e.g. singularity, ridge quality map, orientation field, ridge wavelength map, and skeleton) by human are not considered and used in the proposed matching system, thus the human intervention is remarkably cut down. However, the minutiae extraction is still proceeded manually. In order to avoid the human intervention and achieve high-degree automation, the development of ROI segmentation method and minutiae set-based latent matcher could be another possible solution. Recently, the ROI detector and the minutiae level matching algorithm have been the subject of several studies.

Several approaches have been proposed to address the problem of ROI segmentation. An automated ROI segmentation technique is presented in \cite{Ashtiani08}. For the proposed technique, however, the local orientation and spatial frequency are estimated by using a local pixel intensity projection which is sensitive to the variation of pixel intensity caused by structured noise. \cite{Short11} generates a ideal ``ridge-valley" pattern template and then uses the cross-correlation between a local image patch and the generated template to evaluate the local fingerprint quality. The frequency of ideal ``ridge-valley" pattern template is predefined according to the fixed empirical value, therefore, the generated template is not adaptive to the real local spatial frequency. A total variation (TV) model-based approach is proposed to handle latent fingerprint segmentation task. Therein, the latent fingerprint image is decomposed into cartoon and texture layers and the ROI is detected based on the texture layer by using traditional segmentation methods \cite{Zhang12} \cite{Zhang13}. However, the proposed TV model incorporates the orientation field which is directly calculated from the original poor-quality latent image. Thus, the directional information used in the proposed TV model is not reliable. \cite{Choi12} presents an automated method based on orientation tensor and local ridge frequency to concurrently localize ROI in latent images. However, the local ridge frequency directly estimated from local Fourier analysis on original latent patch is sensitive to the presence of structured noise. \cite{Cao14} proposes a dictionary learning-based segmentation and enhancement method, where the multi-layer ridge structure dictionaries from the coarse level to fine level are separately established by using dictionary learning algorithm. Such approach heavily relies on the learned dictionaries and the training patches are pre-selected from the good-quality rolled fingerprint images. It is demonstrated that a dictionary learned from the target image is preferable (target image means the image currently being processed), since such dictionary can be more adaptive to the target image \cite{Elad06}. However, the ridge structure dictionaries are not learned from the query latent images but from the rolled ones. Therefore, the potentially useful ``ridge-valley" pattern in latent fingerprint images are not utilized but ignored in \cite{Cao14}.

For the reported latent fingerprint matching algorithms, the manual markup of minutiae is regarded as a common practice in latent fingerprint identification cases. This is not only to ensure the matching accuracy, but also to keep the latent matcher in proper working condition. \cite{Paulino11} and \cite{Paulino13} only consider the manually marked minutiae as the sole input for hough transform (HT)-based matcher. Due to the involvement of manually marked minutiae (ground-truth), the proposed matcher could achieve satisfactory matching result. However, the straightforward adoption of the minutiae extracted by automated computer programs (e.g. Verifinger SDK) on low-quality latent images is most likely to result in the poor matching performance. Because of poor quality and overlapping structured noise in latent images, a fair amount of spurious minutiae are possibly yielded via a full-automatic procedure. Consequently, the proposed latent matcher is not robust but vulnerable to the corruption caused by spurious minutiae. \cite{Jain08} and \cite{Jain11} also propose the base-line matching algorithm which only takes the manually marked minutiae as the matcher input. Similar to \cite{Paulino11} and \cite{Paulino13}, the proposed matching approach is also sensitive to the presence of spurious minutiae. Apparently, the robustness and tolerance of minutiae-level matcher for spurious minutiae is therefore a important property and plays a critical role when fulfilling the poor-quality fingerprint matching duties.

The work on robust fingerprint matcher for rolled prints has been proposed. \cite{Tan06} proposes a fingerprint matcher based on genetic algorithm (GA) in order to deal with the significant occlusion and clutter of minutiae caused by low-quality prints. This method achieves good performance when handling low-quality rolled prints. However, the matching performance for latent prints is still unknown. Further, the proposed matcher heavily depends on the local minutiae triangle-based fitness function. Since the local triangle generation based on each triplet of minutiae is computationally intensive, such type of fitness function is too inefficient for GA-based optimization to solve the large size minutiae set matching problem.

In this paper, we propose a robust minutiae set-based matcher embedding with a self-learning module for ROI identification in latent fingerprint images. The proposed latent matcher integrates the following two modules: (i) the dictionary learning (DL)-based ROI segmentation scheme; and (ii) the GA-based minutiae set matching unit. For the DL-based ROI segmentation scheme, the dictionary is firstly learned from the query latent fingerprint image. Then, based on the learned dictionary, the ``ridge-valley" pattern elements (dictionary atoms) can be automatically identified. Further, the sparse representation for the original latent image patches is performed. Finally, depending on the presence or absence of the sparse coefficients that are corresponding to the identified ``ridge-valley" atoms, the foreground (fingerprint region) is segmented. In the GA-based minutiae-level matching unit, the two minutiae sets, one from the segmented ROI in query latent image (obtained via segmentation module) and the other one from the print currently being compared, are aforehand extracted through a normal automated minutiae extraction program like Verifinger SDK which is widely available to the public. Then, according to the affine transformation parameters estimated by GA, the minutiae set alignment between the query latent and the compared print is performed. Further, after aligning two sets of minutiae, the correspondence between the two sets needs to be found. Accordingly, the corresponding minutiae points between the query latent and the compared print could be paired. Finally, the number of matched minutiae is obtained and simply regarded as the matching score.

The main contributions of this paper are summarized as follows:

\begin{enumerate}

  \item To our knowledge, there is no state-of-art latent fingerprint matcher in available public domain. Therefore, in this paper, we introduce a multi-module matcher to cope with the latent fingerprint matching problem. The proposed system is performed in a full-automatic mode. Experimental results based on NIST SD$27$ demonstrate that the proposed matcher with the proposed segmentation module (SM) (say, proposed matcher $+$ proposed SM) can achieve $34.496\%$ penetration rate. The comparative experiments have been conducted to evaluate the effect of the different SMs by using the proposed matcher with and without SM. By designating the proposed matcher without SM as the baseline (say, proposed matcher only), such benchmark penetration rate is $38.159\%$. Based on the benchmark, the proposed matcher with the state-of-art SM such as \cite{Cao14} (say, proposed matcher $+$ SM \cite{Cao14}) only achieves $36.434\%$ penetration rate. The relative penetration rate enhancement percentage for ``proposed matcher $+$ proposed SM" ($9.59\%  = \frac{{\left| {34.496\% {\rm{ - 38}}{\rm{.159\% }}} \right|}}{{{\rm{38}}{\rm{.159\% }}}}$) is at least twice better than that of ``proposed matcher $+$ SM \cite{Cao14}" ($4.52\%  = \frac{{\left| {{\rm{36}}{\rm{.434}}\% {\rm{ - 38}}{\rm{.159\% }}} \right|}}{{{\rm{38}}{\rm{.159\% }}}}$).

  \item The fully automated ROI segmentation module is plug into the proposed latent matcher and is performed as the pre-processing for the subsequent matching task. The proposed SM consists of the following phases: (i) the image structure dictionary learning; (ii) the ``ridge-valley" atom identification; and (iii) the sparse coding and ROI segmentation. Existing method requires to establish a dictionary from the high-quality rolled image patches in advance \cite{Cao14}. Different from such conventional methods, we propose to build up the structure dictionary directly learned from the query latent image. As demonstrated in \cite{Elad06}, a learned dictionary based on the target image can better adapt to the target image. Therefore, the dictionary obtained in the proposed SM is not good quality rolled image patches-determined but query latent image-oriented.

  \item The robust latent matching unit consists of the following stages: (i) the ROI-based minutiae extraction; (ii) the GA-based minutiae set alignment; and (iii) the counting of paired minutiae. Existing method demands to yield the local minutiae descriptors during the iteration of GA optimization \cite{Tan06}. Different from such conventional method, the global topology of the entire minutiae set is directly adopted in the proposed matching unit instead of the local minutiae structure. Accordingly, the proposed matching unit is more robust to the presence of spurious minutiae and more efficient in the matching.

\end{enumerate}

The rest of this paper is organized as follows: in Section \uppercase\expandafter{\romannumeral2}, the details of the proposed matching system are introduced; in Section \uppercase\expandafter{\romannumeral3}, ROI-based minutiae extraction, and latent fingerprint matching experiments are implemented respectively. In the ROI-based minutiae extraction experiment, by adopting the obtained ROI in automated segmentation module, the reduction of spurious minutiae points as well as the preservation of genuine ones are assessed. In latent fingerprint matching experiment, the matching performance of the introduced latent matcher is evaluated; in Section \uppercase\expandafter{\romannumeral4}, the conclusions and on-going research directions are presented.

\section{Proposed Latent Fingerprint Matching System}

In this section, the proposed multi-module matching system for latent fingerprint is proposed, which consists of the two following modules: (i) the dictionary learning-based ROI segmentation scheme; and (ii) the genetic algorithm (GA)-based minutiae set matching unit.
%The following subsections provide the specific description for technical details regarding to the proposed multi-module matcher.

\subsection{Dictionary Learning-Based ROI Segmentation Module}

\subsubsection{Query Latent Image-Based Dictionary Learning}

The dictionary learning procedure is performed based on the query latent fingerprint image instead of the good-quality rolled or plain print images. As suggested in \cite{Elad06}, a learned dictionary based on the target image can better adapt to the target image and more specifically represent the intrinsic signal structure in target image. Compared with other images-based learned dictionary, the target image-based dictionary can more effectively keep the scale consistency for the real signal structures. That is, the signal structure scales in other images might be more or less different from the ones in target image. Therefore, the other images-based dictionary atoms often do not fittingly model the real-scale structures in target image. Accordingly, the image restoration tasks such as denoising and inpainting are conducted depending on the target image-based dictionary rather than the other image-based dictionary or the pre-defined analytic dictionary (e.g. wavelets, curvelets and DCT) \cite{Elad06}. Inspired by the advantage of the target image-based dictionary, the dictionary learned from the query latent image is beneficial to the following ``ridge-valley" atom identification phase.

\begin{figure}
    \centering
    \subfigure[]
    {
        \includegraphics[height=2in]{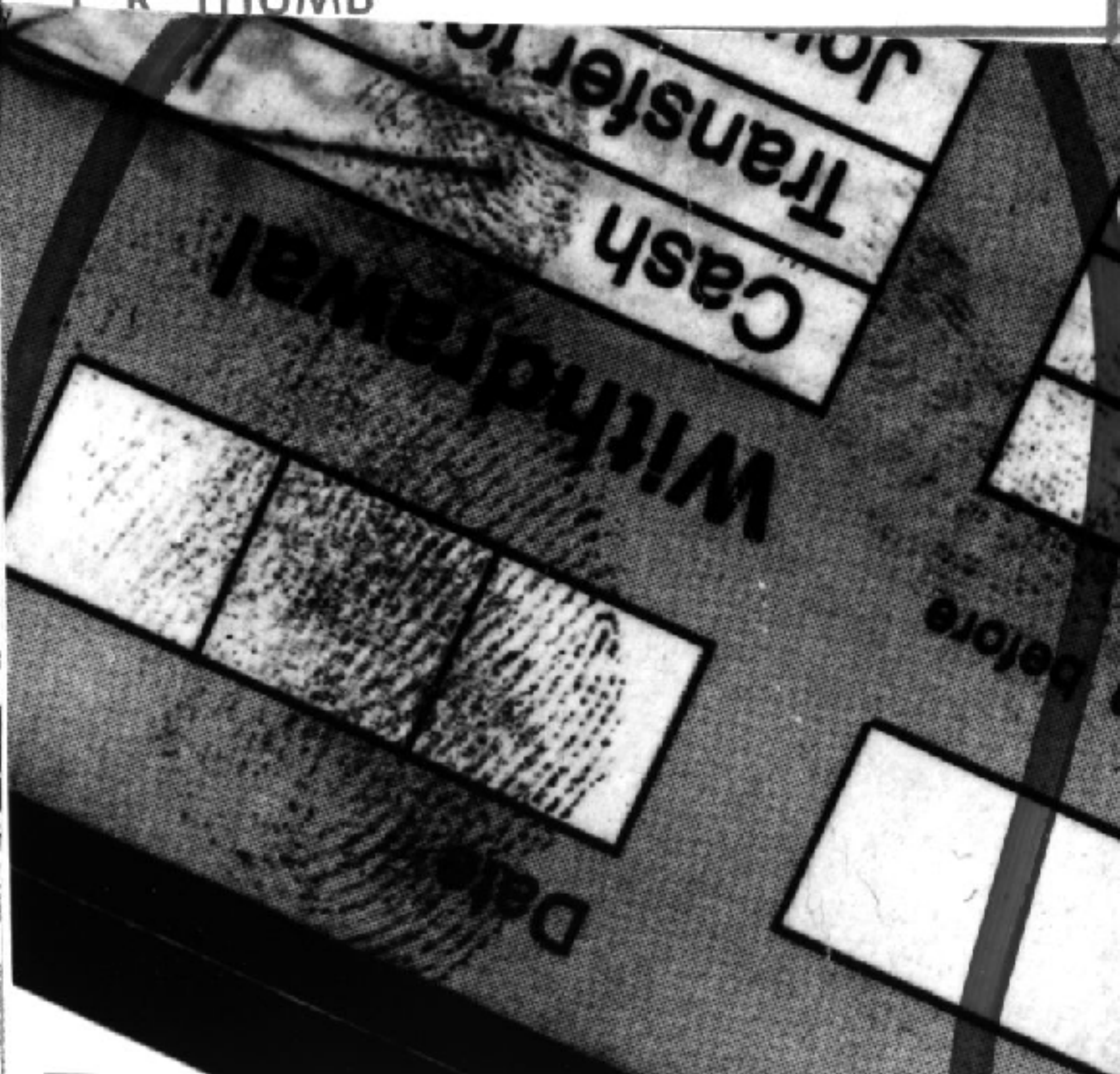}
        \label{show_DL_result:ori_latent}
    }
    \subfigure[]
    {
        \includegraphics[height=2in]{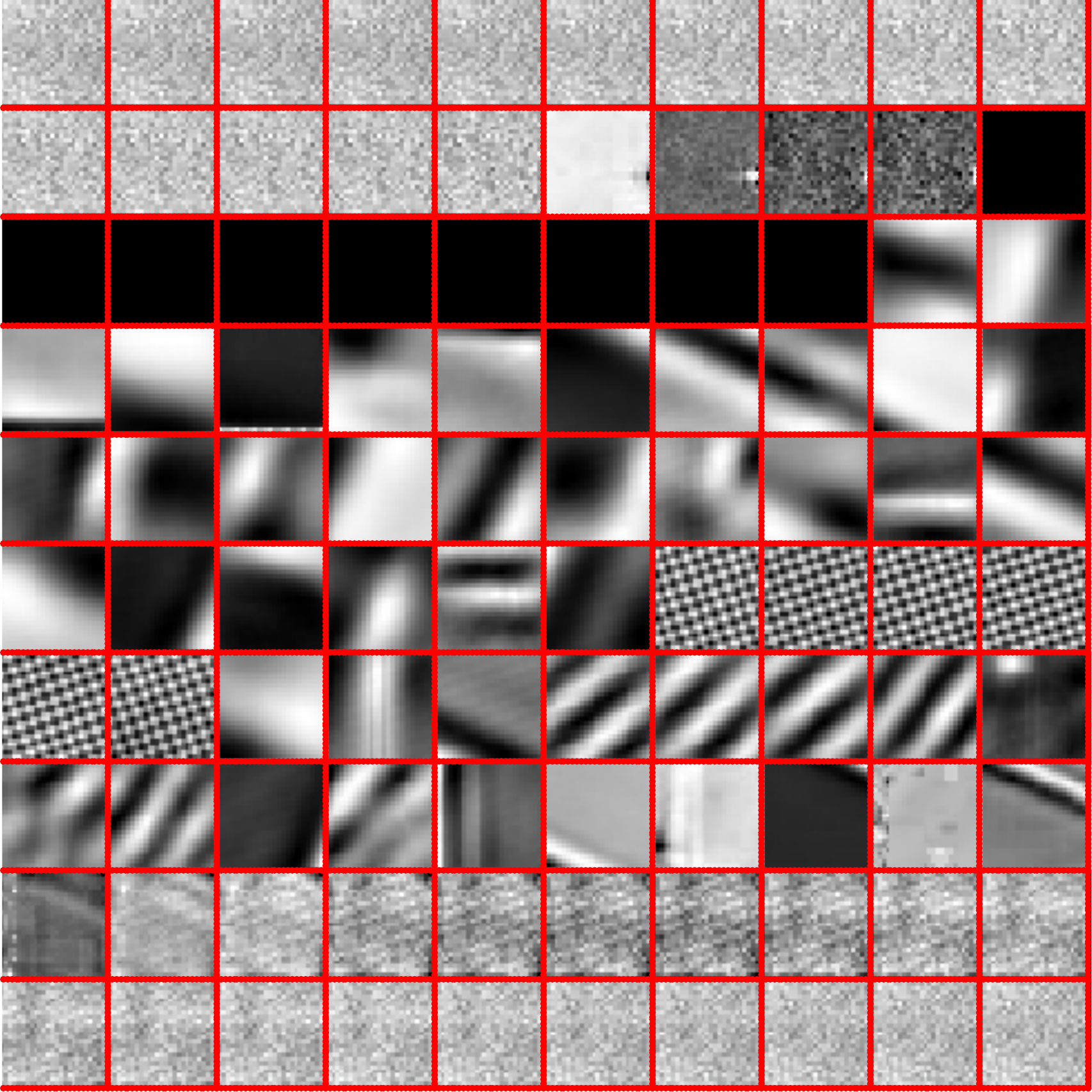}
        \label{show_DL_result:dictionary}
    }
    \caption{The dictionary $D$ is learned directly from the given query latent image by ODL: (a) original query latent image (U224 in NIST SD27); and (b) its learned dictionary $D$ ($N_a = 100$ and $K = 2$).}
    \label{show_DL_result}
\end{figure}

The dictionary learning for query latent image can be mathematically formulated as follows: let $S = \left\{ {{s_i}\left| {i = 1,2,...,N} \right.} \right\}$ as the training set, where $s_i$ is the vector obtained after the vectorization of latent image block $p_i$ with size $w \times w$ and $N$ is the number of selected image blocks in latent fingerprint image. The purpose of dictionary learning is to establish a numerical dictionary $D$ with size ${N_s} \times {N_a}$ based on the training set $S$ ($N_s$ denotes the atom vector dimension where $N_s = w \times w$ and $N_a$ stands for the atom number). Such established dictionary $D$ can effectively represent the vectorized image block $s_i$ in a sparse way. In order to obtain dictionary $D$, the following optimization problem needs to be solved

\begin{equation}
\mathop {\min }\limits_{D,\Gamma } \left\| {S - D\Gamma } \right\|_F^2\begin{array}{*{20}{c}}
{}
\end{array}s.t.\begin{array}{*{20}{c}}
{}
\end{array}\forall i,\begin{array}{*{20}{c}}
{}
\end{array}{\left\| {{\gamma _i}} \right\|_0} \le K
\label{ori_opt_pro}
\end{equation}

where $\Gamma$ is the sparse coefficient matrix with size ${N_a} \times N$. $\gamma_i$ is the $i^{th}$ column vector in sparse coefficient matrix $\Gamma$ and is also corresponding to the $i^{th}$ vectorized image block $s_i$. ${\left\|  \cdot  \right\|_F}$ and ${\left\|  \cdot  \right\|_0}$ denote the Frobenius norm and $l_0$ norm respectively. ${\left\| {{\gamma _i}} \right\|_0}$ is equal to the number of nonzero elements in vector $\gamma_i$. Therefore, in order to obtain the sparse vector $\gamma_i$, the regulator ${\left\| {{\gamma _i}} \right\|_0} \le K$ controls that the nonzero elements in vector $\gamma_i$ should be less than the sparsity parameter $K$. Although $l_0$ is the limit of $l_p$ when $p$ approaches zero, it is not a true norm (unlike the $l_1$ norm which has all properties of a true norm) and also leads to the NP-hard problem. In order to avoid the NP-hard problem, the above $l_0$ norm-based regulator ${\left\| {{\gamma _i}} \right\|_0} \le K$ in Equation (\ref{ori_opt_pro}) can be replaced by the convex $l_1$ norm-based regulator $\min \left( {{{\left\| {{\gamma _i}} \right\|}_1}} \right)$. Accordingly, Equation (\ref{ori_opt_pro}) is updated by the following approximate form

\begin{equation}
\mathop {\min }\limits_{D,\Gamma } \left( {\left\| {S - D\Gamma } \right\|_2^2 + \lambda \sum\limits_{i = 1}^N {{{\left\| {{\gamma _i}} \right\|}_1}} } \right)
\label{updated_opt_pro}
\end{equation}

where Frobenius norm ${\left\|  \cdot  \right\|_F}$ has been embodied by $l_2$ norm and $\lambda$ is the Lagrange coefficient to balance the data fitting and sparsity level. For solving the optimization problem in Equation (\ref{updated_opt_pro}), the sparse coefficient matrix $\Gamma$ and the dictionary $D$ are alternatively updated as follows: (i) keeping $D$ fixed, compute the sparse coefficient matrix $\Gamma$; then (ii) keeping $\Gamma$ fixed, update the dictionary $D$. The above two steps are iteratively repeated until the convergence. In order to learn a dictionary $D$, one can apply any available dictionary learning technique such as MOD \cite{Engan2000}, KSVD \cite{Elad06} and ODL \cite{Mairal09}. As reported in \cite{Mairal09}, an alternate optimization of sparse coding and dictionary update is performed based on a subset of the training data. Such subset continues to be augmented with a new coming training sample. Based on the outcome of the previous iteration, the same alternate optimization is executed again for the new coming training data. ODL iteratively repeats until all training samples have been used. As demonstrated by the experiments in \cite{Mairal09}, ODL is more faster than MOD and KSVD. Because the large size of training set is produced based on the dense training set pick-up strategy and the whole training set has to be used by MOD and KSVD at each iteration, both MOD and KSVD are computationally expensive. Therefore, instead of MOD and KSVD, ODL is applied to learn the dictionary for query latent fingerprint image due to its more efficient learning mechanism and lower computational complexity. The dictionary $D$ learned from the given query latent image by using ODL is shown in Figure \ref{show_DL_result}.

%------------------- show original latent / dictionary -------------------
%------------------- show original latent / dictionary -------------------

\subsubsection{Ridge-Valley Atom Identification}

The atoms with ``ridge-valley" pattern need to be distinguished from the ones which are not structured as ``ridge-valley" pattern. The detection or modeling of ``ridge-valley" pattern is not a new subject and the previous works have been reported in \cite{Ashtiani08} and \cite{Short11}, however, it is still a challenging task. Although ``ridge-valley" pattern is one type of intrinsic signal structure in latent fingerprint image, such pattern is usually mingled with other types of structured noise (e.g. arch, line, character, stain, speckle and motif) and even vague or disconnected due to the wet or dry prints. Accordingly, the direct detection or straightforward modeling for such pattern based on original latent image patch is inaccurate. In contrast to the original patch-level detection, the atom-level detection is easier and more reliable. That is, the dictionary learning procedure is not only for the signal structural decomposition but also for the signal structure refinement. After the completion of dictionary learning, the intrinsic signal structures at atom level becomes more salient and easier to be distinguished. Because the identified ``ridge-valley" atoms play a crucial role in the subsequent sparse coefficient-based ROI segmentation phase, the development of a fully automated identification approach is necessary.

In this study, a local Fourier analysis and cross correlation-based method is proposed, which consists of the following steps:

%------------------- show "ridge-valley" atom -------------------
\begin{figure}
    \centering
    \subfigure[]
    {
        \includegraphics[height=1.5in]{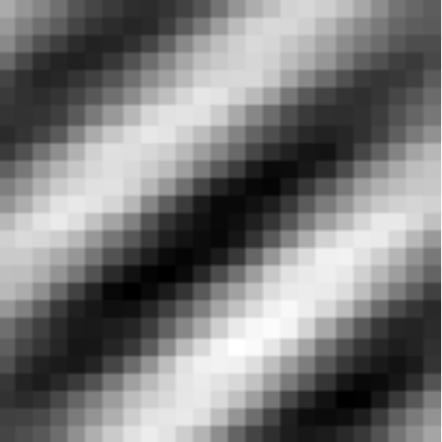}
        \label{show_ridge_valley_atom_identification:ridge_valley_atom}
    }
    \subfigure[]
    {
        \includegraphics[height=1.5in]{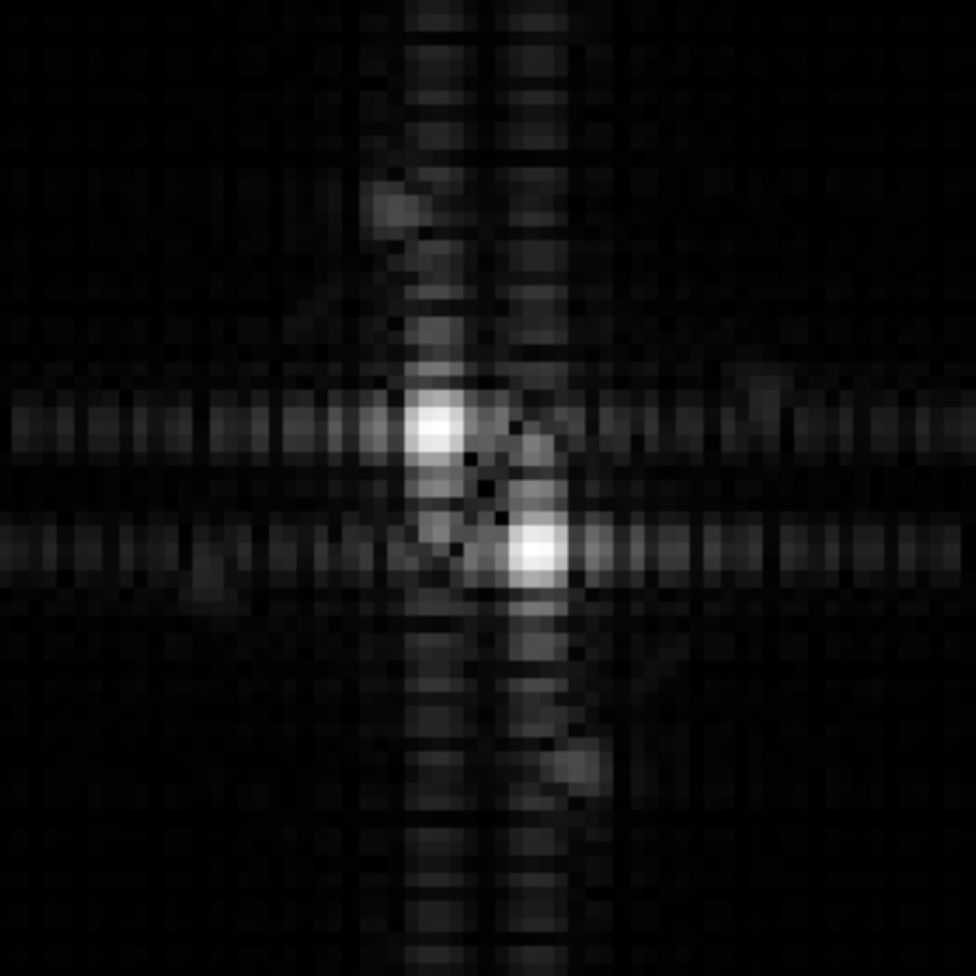}
        \label{show_ridge_valley_atom_identification:ridge_valley_atom_fft}
    }
    \subfigure[]
    {
        \includegraphics[height=1.5in]{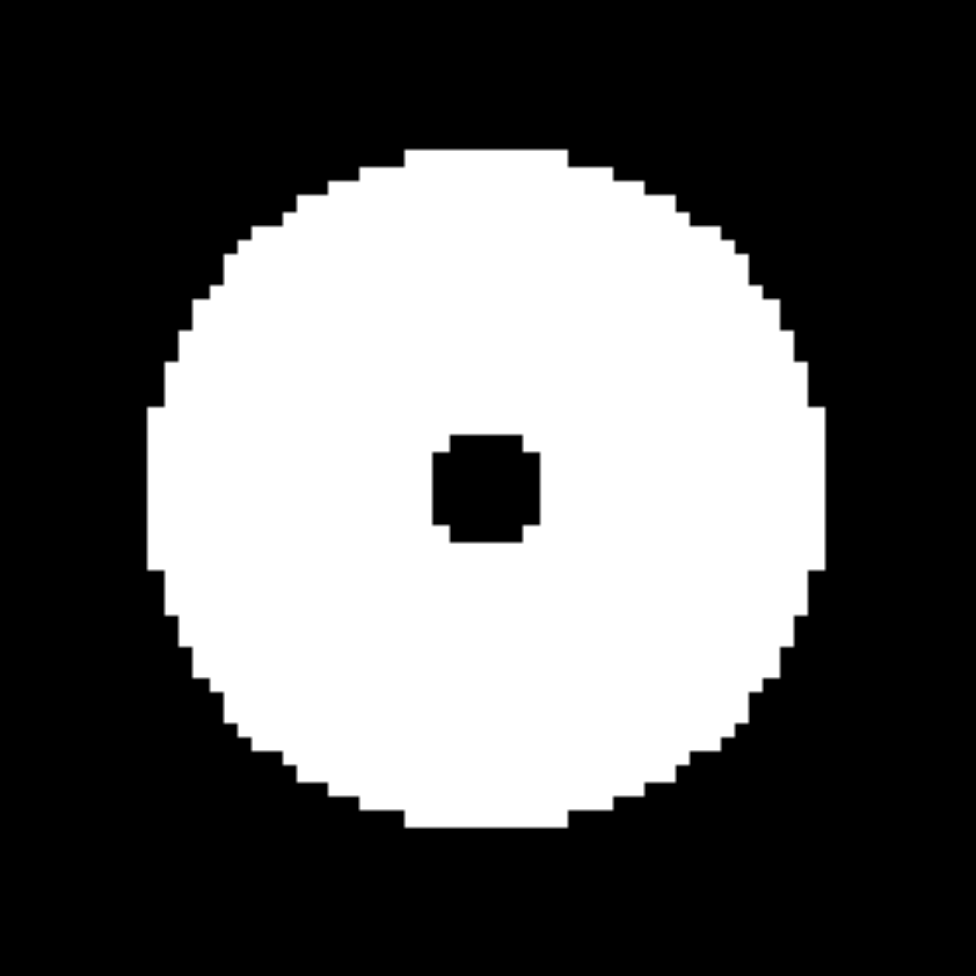}
        \label{show_ridge_valley_atom_identification:ridge_valley_band}
    }
    \subfigure[]
    {
        \includegraphics[height=1.5in]{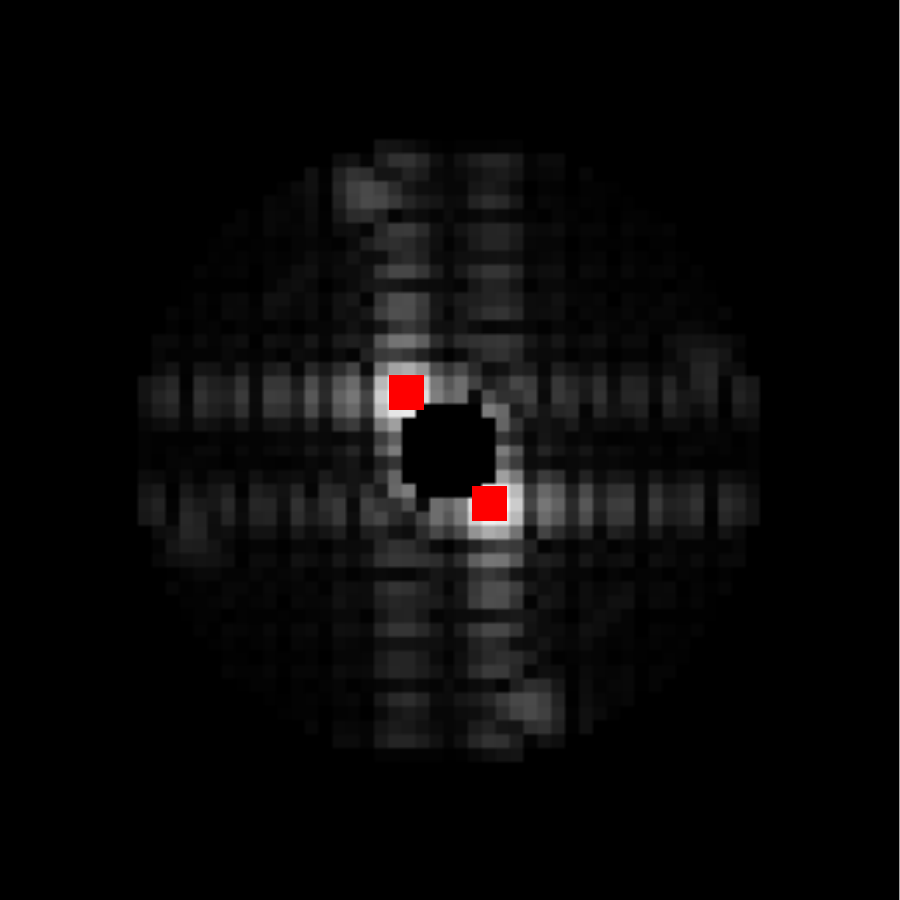}
        \label{show_ridge_valley_atom_identification:ridge_valley_constrained_fft}
    }
    \subfigure[]
    {
        \includegraphics[height=1.5in]{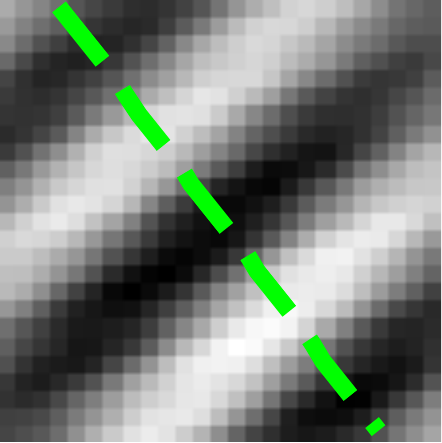}
        \label{show_ridge_valley_atom_identification:ridge_valley_orientation}
    }
    \subfigure[]
    {
        \includegraphics[height=1.5in]{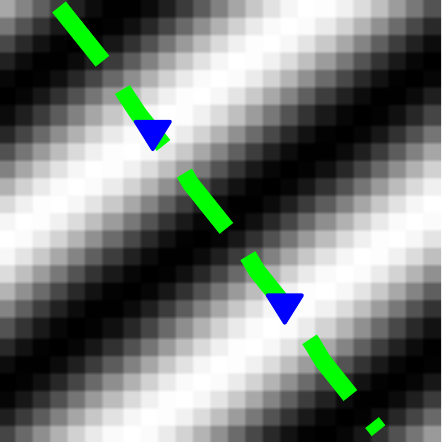}
        \label{show_ridge_valley_atom_identification:ridge_valley_reconstruction}
    }
    \subfigure[]
    {
        \includegraphics[height=2in]{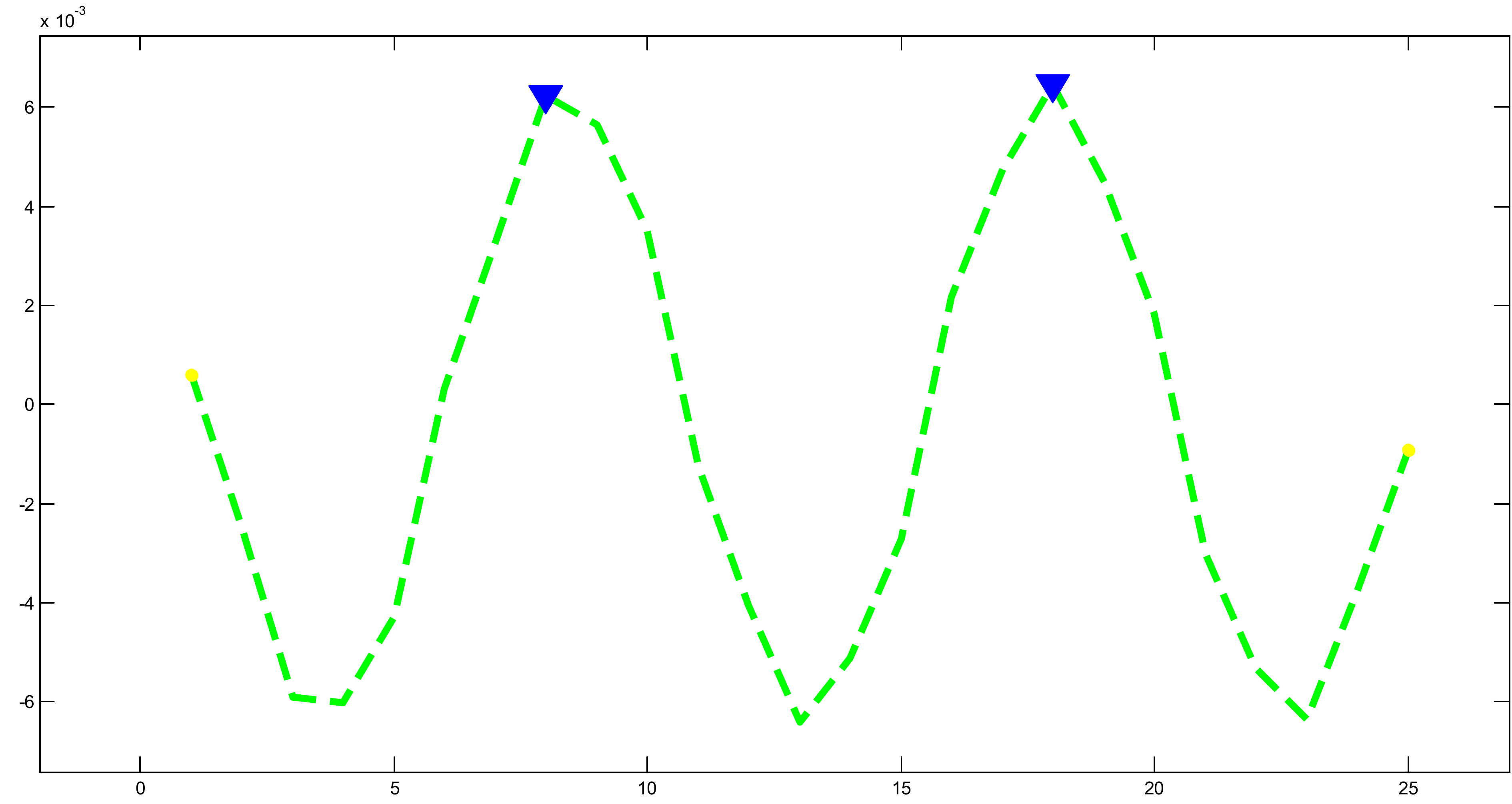}
        \label{show_ridge_valley_atom_identification:ridge_valley_1d_sine_wave}
    }
    \caption{The illustration of the details during the automated ``ridge-valley" atom identification procedure: (a) the ``ridge-valley" atom patch (${p_{{d_{66}}}}$ in Figure \ref{show_DL_result:dictionary}); (b) the spectral magnitude map $\left| {DFT\left( {{p_{{d_{66}}}}} \right)} \right|$; (c) the frequency bandpass filter corresponding to the ridge period range $[3 \ pixels, 20 \ pixels]$; (d) the constrained spectral magnitude map depending on (c); (e) the calculated orientation $o_{d_{66}}$; (f) the reconstructed ``ridge-valley" pattern $u_{d_{66}}$; and (g) the 1D sinusoidal-shaped wave modeled based on the pixel intensities along the green dashed line indicated in (f).}
    \label{show_ridge_valley_atom_identification}
\end{figure}
%------------------- show "ridge-valley" atom -------------------

%------------------- show "ridge-valley" and "non-ridge-valley" atom -------------------
\begin{figure}
    \centering
    \subfigure[]
    {
        \includegraphics[height=1.5in]{ridge_valley_atom.pdf}
        \label{show_different_atoms:ridge_valley_pattern_1}
    }
    \subfigure[]
    {
        \includegraphics[height=1.5in]{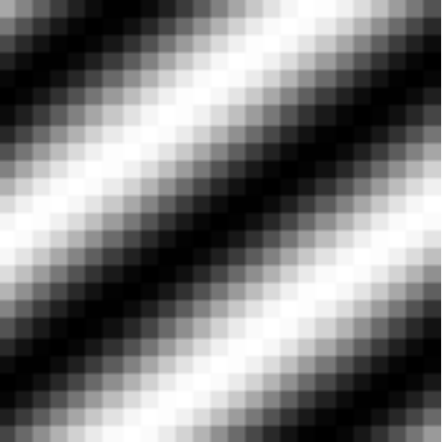}
        \label{show_different_atoms:reconstructed_ridge_valley_pattern_1}
    }
    \subfigure[]
    {
        \includegraphics[height=1in]{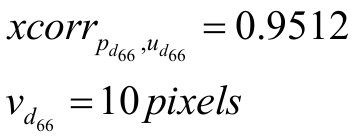}
        \label{show_different_atoms:xcorr_1}
    }
    \subfigure[]
    {
        \includegraphics[height=1.5in]{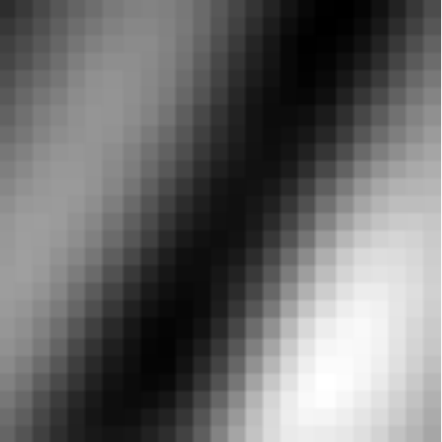}
        \label{show_different_atoms:non_ridge_valley_pattern_2}
    }
    \subfigure[]
    {
        \includegraphics[height=1.5in]{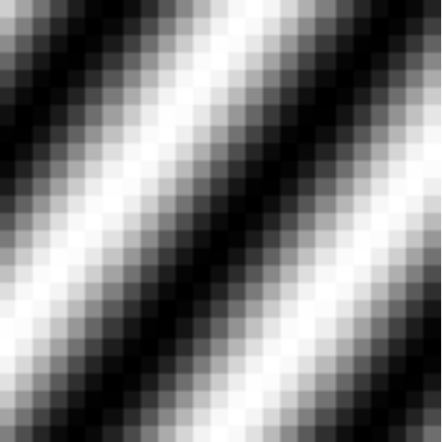}
        \label{show_different_atoms:reconstructed_ridge_valley_pattern_2}
    }
    \subfigure[]
    {
        \includegraphics[height=1in]{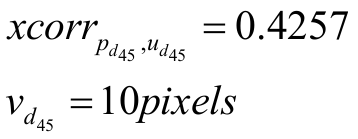}
        \label{show_different_atoms:xcorr_2}
    }
    \subfigure[]
    {
        \includegraphics[height=1.5in]{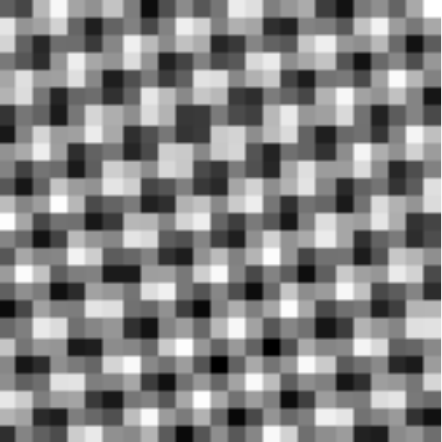}
        \label{show_different_atoms:non_ridge_valley_pattern_3}
    }
    \subfigure[]
    {
        \includegraphics[height=1.5in]{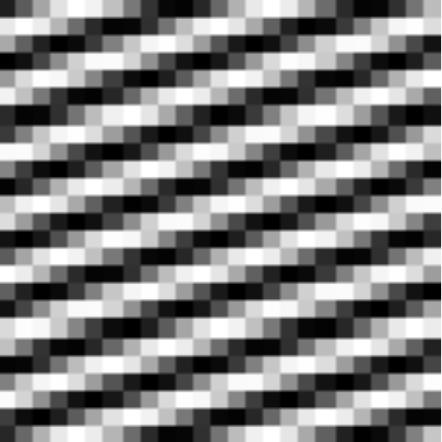}
        \label{show_different_atoms:reconstructed_ridge_valley_pattern_3}
    }
    \subfigure[]
    {
        \includegraphics[height=1in]{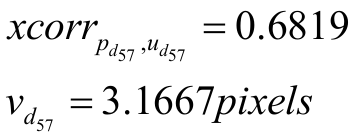}
        \label{show_different_atoms:xcorr_3}
    }
    \caption{The examples of atom-based identification for various atoms: (a) the ``ridge-valley" atom (${p_{{d_{66}}}}$ in Figure \ref{show_DL_result:dictionary}); (b) and (c) are the reconstructed $u_{d_{66}}$, the calculated $xcor{r_{{p_{{d_{66}}}},{u_{{d_{66}}}}}}$ and ${v_{{d_{66}}}}$ for (a) respectively; (d) the line-like atom (${p_{{d_{45}}}}$ in Figure \ref{show_DL_result:dictionary}); (e) and (f) are the reconstructed $u_{d_{45}}$, the calculated $xcor{r_{{p_{{d_{45}}}},{u_{{d_{45}}}}}}$ and ${v_{{d_{45}}}}$ for (d) respectively; (g) the periodical speckle atom (${p_{{d_{57}}}}$ in Figure \ref{show_DL_result:dictionary}); (h) and (i) are the reconstructed $u_{d_{57}}$, the calculated $xcor{r_{{p_{{d_{57}}}},{u_{{d_{57}}}}}}$ and ${v_{{d_{57}}}}$ for (g) respectively.}
    \label{show_different_atoms}
\end{figure}
%------------------- show "ridge-valley" and "non-ridge-valley" atom -------------------

\begin{itemize}

  \item {\it Step 1:} Given a learned dictionary $D = \left\{ {{d_k}\left| {k = 1,2,...,{N_a}} \right.} \right\}$, each single atom vector $d_k$ with dimension $N_s$ needs to be converted to the atom patch ${p_{{d_k}}}$ with size $w \times w$ ($N_s = w \times w$) (shown in Figure \ref{show_ridge_valley_atom_identification:ridge_valley_atom});

  \item {\it Step 2:} Input a atom patch ${p_{{d_k}}}$, the 2D Discrete Fourier Transform (2D DFT) is applied to ${p_{{d_k}}}$, then the spectrum $DFT\left( {{p_{{d_k}}}} \right)$ in frequency domain is obtained (shown in Figure \ref{show_ridge_valley_atom_identification:ridge_valley_atom_fft});

  \item {\it Step 3:} In the spectral magnitude map $\left| {DFT\left( {{p_{{d_k}}}} \right)} \right|$ (rearranged by moving the zero-frequency component to the center), the paired highest magnitude points are detected within the frequency range corresponding to the ridge period range $[3 \ pixels, 20 \ pixels]$. Such ridge period range can cover most image underlying structures from low spatial frequency to high spatial frequency. The bandpass constraint in frequency domain corresponding to the ridge period range $[3 \ pixels, 20\ pixels]$ in spatial domain is shown in Figure \ref{show_ridge_valley_atom_identification:ridge_valley_band}. Such frequency bandpass filter masks over the spectral magnitude map to yield the constrained spectral magnitude map, and the highest magnitude points can be detected and shown in Figure \ref{show_ridge_valley_atom_identification:ridge_valley_constrained_fft} (marked by red squares).

  \item {\it Step 4:} Based on the coordinate of paired points detected in {\it Step 3}, the orientation ${o_{{d_k}}}$ for the currently processed atom patch ${p_{{d_k}}}$ is calculated (for ``ridge-valley" atom, the orientation ${o_{{d_k}}}$ is not along but across the ridge, as indicated by the green dashed line in Figure \ref{show_ridge_valley_atom_identification:ridge_valley_orientation});

  \item {\it Step 5:} Reconstruct the ``ridge-valley" pattern ${u}_{{d_k}}$ according to the magnitude and phase corresponding to the detected points in constrained spectral magnitude map (the reconstructed ${u}_{{d_k}}$ is illustrated in Figure \ref{show_ridge_valley_atom_identification:ridge_valley_reconstruction}), then calculate the cross-correlation value $xcor{r_{{p_{{d_k}}},{u_{{d_k}}}}}$ between the atom patch ${p_{{d_k}}}$ and the reconstructed ``ridge-valley" pattern ${u}_{{d_k}}$ according to Equation (\ref{x_corr}).

    \begin{equation}
    xcor{r_{{p_{{d_k}}},{u_{{d_k}}}}} = \frac{{\sum\limits_{x,y} {\left( {\alpha  \cdot \beta } \right)} }}{{\sqrt {\sum\limits_{x,y} {\left( {{\alpha ^2}} \right)}  \cdot \sum\limits_{x,y} {\left( {{\beta ^2}} \right)} } }}
    \label{x_corr}
    \end{equation}

      where $\alpha  = {f_{{p_{{d_k}}}}}\left( {x,y} \right) - {{\bar f}_{{p_{{d_k}}},{u_{{d_k}}}}}$ and $\beta  = {f_{{u_{{d_k}}}}}\left( {x - a,y - b} \right) - {{\bar f}_{{u_{{d_k}}}}}$. ${{f_{{p_{{d_k}}}}}\left( {x,y} \right)}$ and ${f_{{u_{{d_k}}}}}\left( {x,y} \right)$ denote the atom image ${p_{{d_k}}}$ and the reconstructed pattern ${u}_{{d_k}}$ respectively. ${{\bar f}_{{p_{{d_k}}},{u_{{d_k}}}}}$ stands for the mean of ${{f_{{p_{{d_k}}}}}\left( {x,y} \right)}$ when overlapping with ${f_{{u_{{d_k}}}}}\left( {x,y} \right)$. ${{{\bar f}_{{u_{{d_k}}}}}}$ is the mean of the entire reconstructed pattern image ${u}_{{d_k}}$. $a$ and $b$ denote the offsets along $x-$ and $y-$ axis respectively. Equation (\ref{x_corr}) indicates the pattern similarity between atom patch ${p_{{d_k}}}$ and ``ridge-valley" pattern ${u}_{{d_k}}$. The higher cross-correlation value is, the atom patch ${p_{{d_k}}}$ is more likely to represent the ``ridge-valley" pattern.

%      \newcounter{mytempeqncnt}
%      \begin{figure*}[!t]
%      \normalsize
%      \setcounter{mytempeqncnt}{\value{equation}}
%      \setcounter{equation}{11}
%      \begin{equation}
%      \label{x_corr}
%            xcor{r_{{p_{{d_k}}},{u_{{d_k}}}}} = \frac{{\sum\limits_{x,y} {\left[ {{f_{{p_{{d_k}}}}}\left( {x,y} \right) - {{\bar f}_{{p_{{d_k}}},{u_{{d_k}}}}}} \right] \cdot \left[ {{f_{{u_{{d_k}}}}}\left( {x - a,y - b} \right) - {{\bar f}_{{u_{{d_k}}}}}} \right]} }}{{\sqrt {\sum\limits_{x,y} {{{\left[ {{f_{{p_{{d_k}}}}}\left( {x,y} \right) - {{\bar f}_{{p_{{d_k}}},{u_{{d_k}}}}}} \right]}^2} \cdot \sum\limits_{x,y} {{{\left[ {{f_{{u_{{d_k}}}}}\left( {x - a,y - b} \right) - {{\bar f}_{{u_{{d_k}}}}}} \right]}^2}} } } }}
%      \end{equation}
%      \setcounter{equation}{\value{mytempeqncnt}}
%      \hrulefill
%      \vspace*{4pt}
%      \end{figure*}

  \item {\it Step 6:} For the reconstructed ``ridge-valley" pattern ${u}_{{d_k}}$, the pixel intensities along the orientation ${o_{{d_k}}}$ are acquired to model a 1D sinusoidal-shaped wave, then the wave peaks are detected to calculate the spatial ridge period ${v_{{d_k}}}$. The pixels along ${o_{{d_k}}}$ are highlighted by the green dashed line in Figure \ref{show_ridge_valley_atom_identification:ridge_valley_reconstruction}, and the modeled 1D sinusoidal-shaped wave is shown in Figure \ref{show_ridge_valley_atom_identification:ridge_valley_1d_sine_wave} where the blue triangles indicate the detected wave peaks;

  \item {\it Step 7:} Check whether the following two criterions are concurrently satisfied: (i) $xcor{r_{{p_{{d_k}}},{u_{{d_k}}}}} \ge T{h_{xcorr}}$ ($T{h_{xcorr}}$ is the threshold to determine the pattern similarity); and (ii) ${v_{{d_k}}} \in [5.3 \ pixels, 12.8 \ pixels]$ ($[5.3 \ pixels, 12.8 \ pixels]$ is suggested in \cite{Choi12}. In the proposed atom identification procedure, the broad range $[3 \ pixels, 20 \ pixels]$ is firstly adopted to involve more candidate atoms, then the narrow range $[5.3 \ pixels, 12.8 \ pixels]$ is applied to filter out the unqualified atoms). If both satisfied, the currently processed atom patch ${p_{{d_k}}}$ is regarded as the ``ridge-valley" atom. Otherwise, label ${p_{{d_k}}}$ as the ``non ridge-valley" atom. The judgement depending on the single criterion is too limited to make a correct decision on atom level, therefore both criterions need to be simultaneously satisfied. As demonstrated in Figure \ref{show_different_atoms}, the ``non ridge-valley" atoms either satisfies $xcor{r_{{p_{{d_k}}},{u_{{d_k}}}}} \ge T{h_{xcorr}}$ (here $T{h_{xcorr}}=0.6$) or meets ${v_{{d_k}}} \in [5.3 \ pixels, 12.8 \ pixels]$, however, they are not the ``ridge-valley" atoms indeed;

  \item {\it Step 8:} Check whether all the atoms in $D$ have been judged. If not, go back to {\it Step 2}. Otherwise, terminate the ``ridge-valley" atom identification procedure and output all the atom labels. An example of the ``ridge-valley" atoms identification for all the atoms in learned dictionary is illustrated in Figure \ref{show_all_identified_atoms}.

\end{itemize}

%------------------- show all the identified atoms -------------------
\begin{figure}
    \centering
    \subfigure[]
    {
        \includegraphics[height=2in]{DL_dictionary.pdf}
        \label{show_all_identified_atoms:DL_dictionary}
    }
    \subfigure[]
    {
        \includegraphics[height=2in]{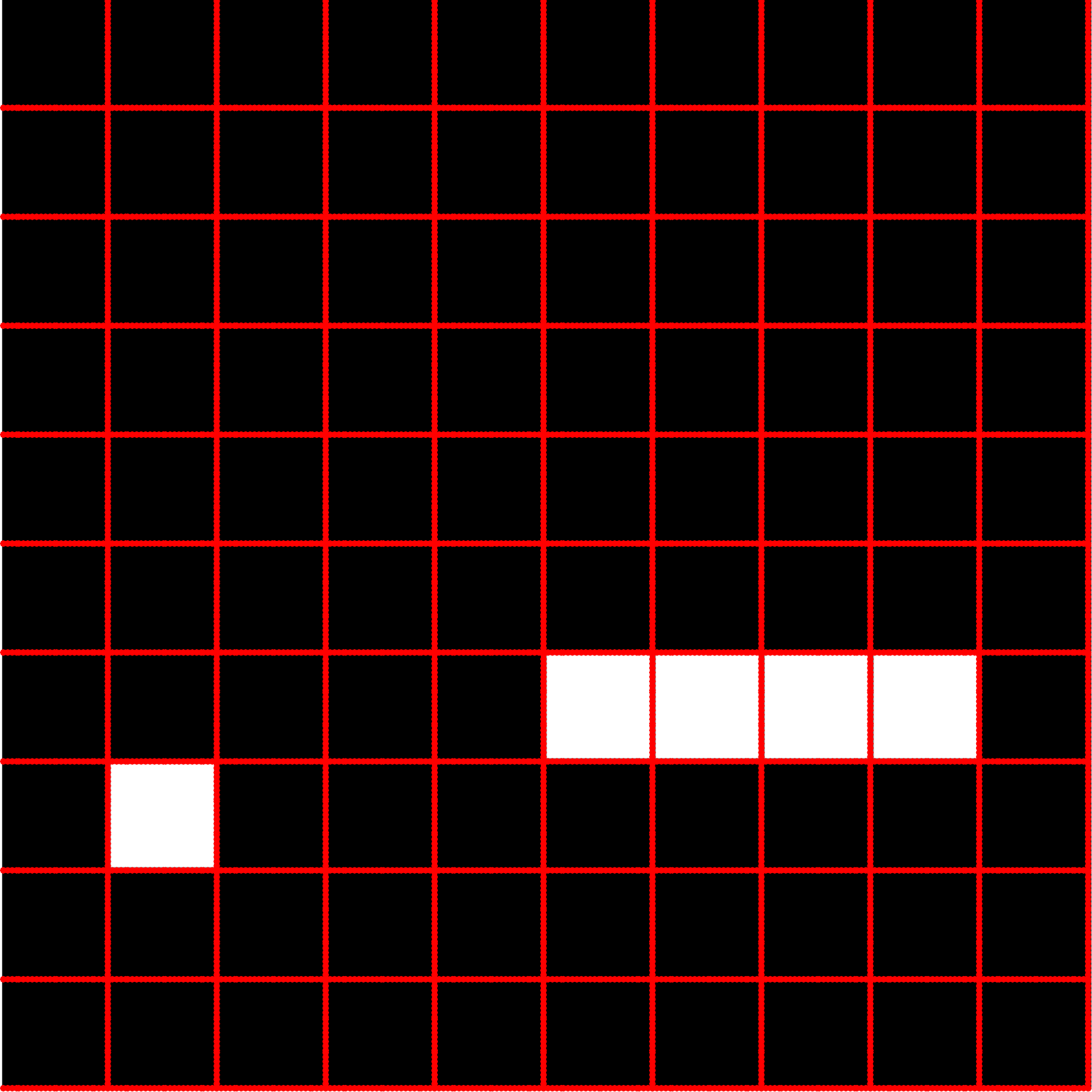}
        \label{show_all_identified_atoms:identified_ridge_valley_atoms}
    }
    \caption{The example of the identification results obtained by using the proposed automated ``ridge-valley" atom identification procedure: (a) the learned dictionary $D$ in Figure \ref{show_DL_result:dictionary}; and (b) the identified ``ridge-valley" atoms. The white patches indicate that the original atoms with the same patch-wise coordinates in (a) are ``ridge-valley" ones, while the black patches label that the corresponding atoms in (a) are ``non-ridge-valley" ones.}
    \label{show_all_identified_atoms}
\end{figure}
%------------------- show all the identified atoms -------------------

\subsubsection{Sparse Coefficient-Based ROI Segmentation}

The sparse coefficients by projecting the image blocks in query latent image onto the learned dictionary are utilized to determine whether the original image blocks belong to the foreground. To be specific, the sparse projection from every single latent image block to the learned dictionary yields the sparse coefficient vector. Given a single latent image block, the elements inside its sparse coefficient vector are quite different. That is, the most elements are zero while few are nonzero. The nonzero elements in sparse coefficient vector measures the similarity between such latent image block and the corresponding atoms. Among the nonzero elements, the higher one indicates the corresponding atom is more similar to such image block, while the lower one indicates the lower similarity. As a latent image block with ``ridge-valley" pattern is considered as one part of real fingerprint, the latent image block whose highest nonzero element in sparse coefficient vector corresponds to the ``ridge-valley" atom is regarded as the subset of foreground. Otherwise, the image block whose highest nonzero element in sparse coefficient vector corresponds to the ``non-ridge-valley" atom is regarded as the background. Motivated by the presence or absence of highest nonzero sparse coefficients corresponding to the identified ``ridge-valley" atoms in preceding stage, the ROI segmentation phase could be smoothly proceeded.

The generation of sparse coefficients can be mathematically formulated as the following equation

\begin{equation}
{{\tilde s}_i} = \gamma _i^{\left( 1 \right)}{d_1} + \gamma _i^{\left( 2 \right)}{d_2} + \gamma _i^{\left( 3 \right)}{d_3} + ... + \gamma _i^{\left( {{N_a}} \right)}{d_{{N_a}}} = \sum\limits_{k = 1}^{{N_a}} {\gamma _i^{\left( k \right)}{d_k}}
\label{sparse_rep}
\end{equation}

where ${{\tilde s}_i}$ is the approximation of the given signal vector $s_i$ ($s_i$ is obtained after the vectorization for the latent image block $p_i$). ${\gamma _i^{\left( k \right)}}$ is the element inside the sparse coefficient vector ${\gamma _i} = {\left[ {\gamma _i^{\left( 1 \right)},\gamma _i^{\left( 2 \right)},\gamma _i^{\left( 3 \right)},...,\gamma _i^{\left( k \right)},...,\gamma _i^{\left( {{N_a}} \right)}} \right]^T}$. Due to the sparsity constraint in Equation (\ref{ori_opt_pro}), most atoms in dictionary $D$ are not selected for the representation of $s_i$, but only a small number of atoms are adopted. Accordingly, most elements in vector $\gamma _i$ are zero while few are nonzero.

In this study, the nonzero elements inside the sparse coefficient vector $\gamma_i$ are used for ROI segmentation. Such sparse coefficient-based ROI segmentation stage consists of the following steps:

%------------------- show foreground / background latent block -------------------
\begin{figure}
    \centering
    \subfigure[]
    {
        \includegraphics[height=1.5in]{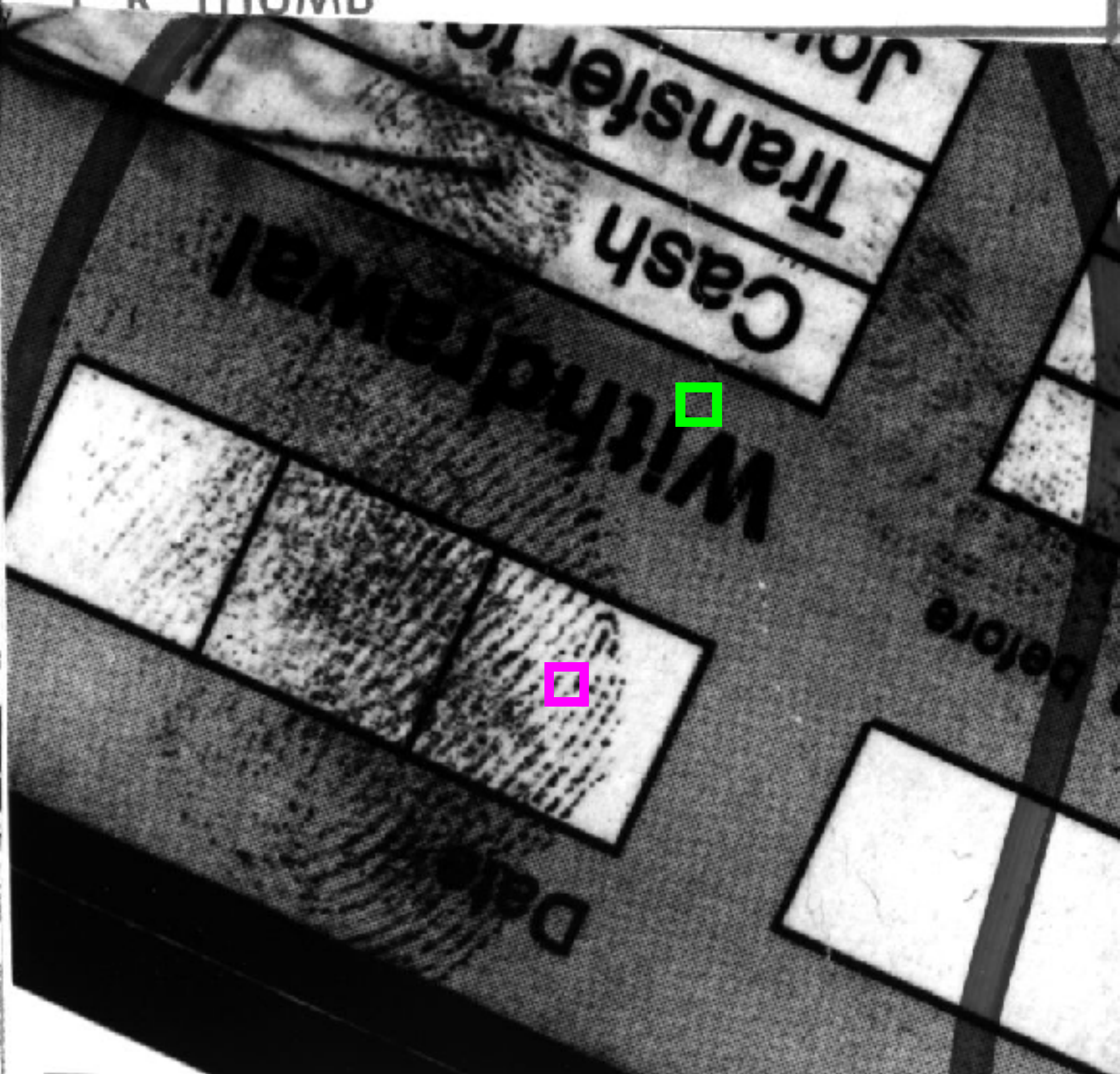}
        \label{show_sparse_coefficient_seg:overview}
    }
    \\
    \subfigure[]
    {
        \includegraphics[height=1.5in]{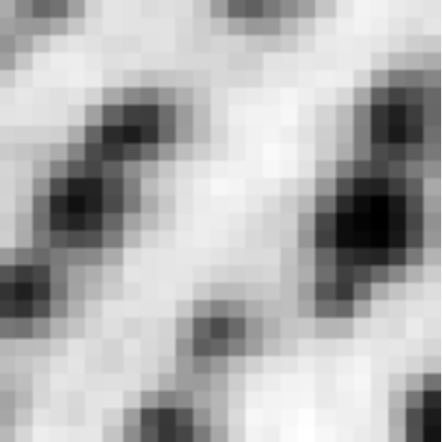}
        \label{show_sparse_coefficient_seg:fore_patch}
    }
    \subfigure[]
    {
        \includegraphics[height=1.5in]{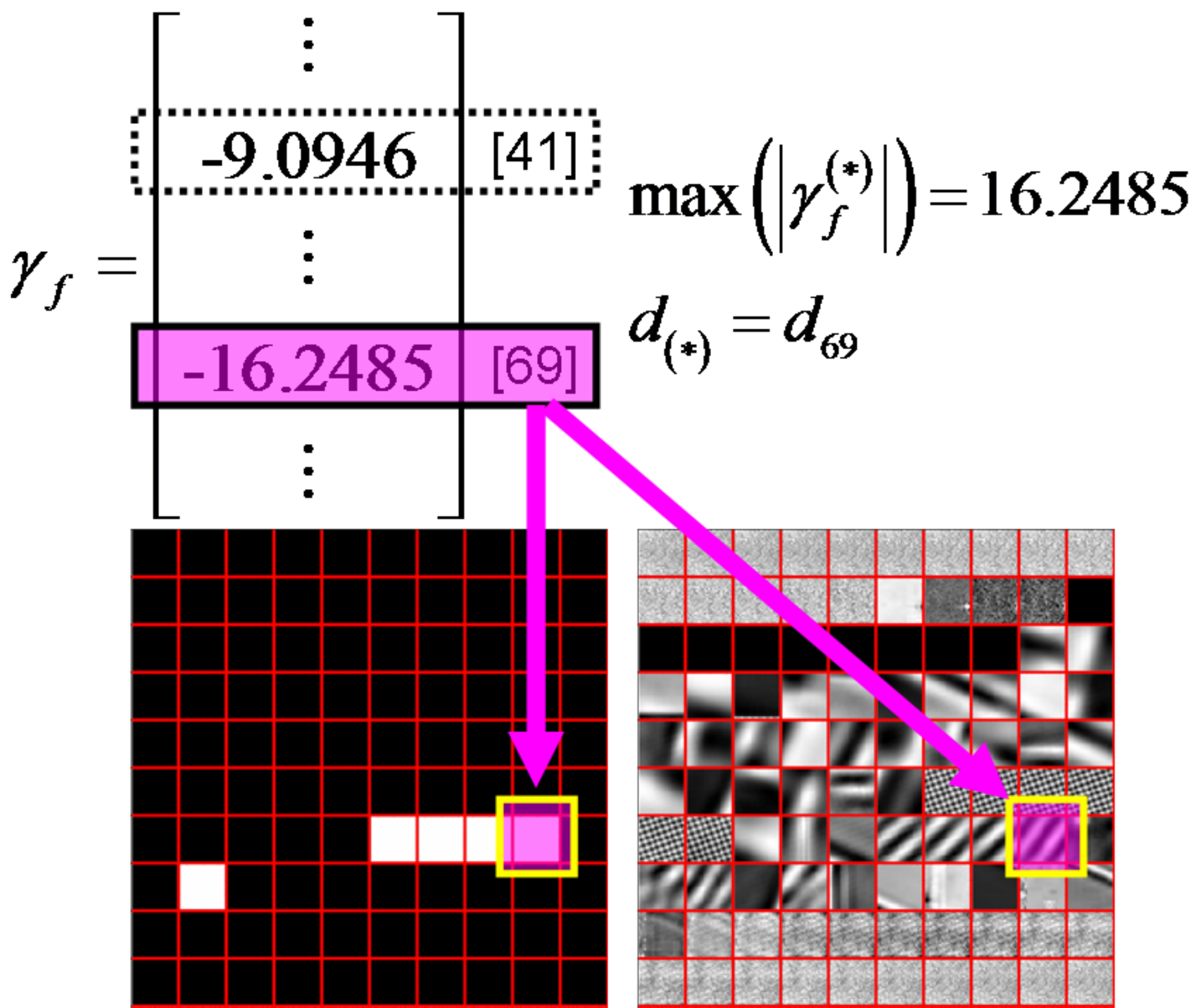}
        \label{show_sparse_coefficient_seg:fore_sparse_vec}
    }
    \\
    \subfigure[]
    {
        \includegraphics[height=1.5in]{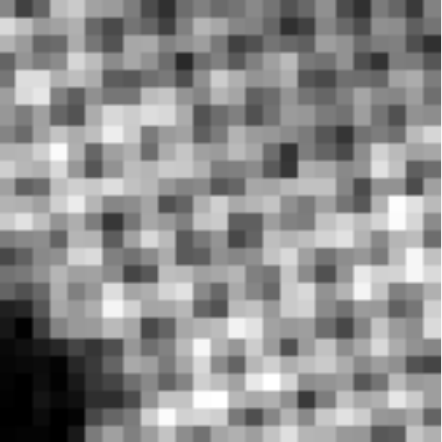}
        \label{show_sparse_coefficient_seg:back_patch}
    }
    \subfigure[]
    {
        \includegraphics[height=1.5in]{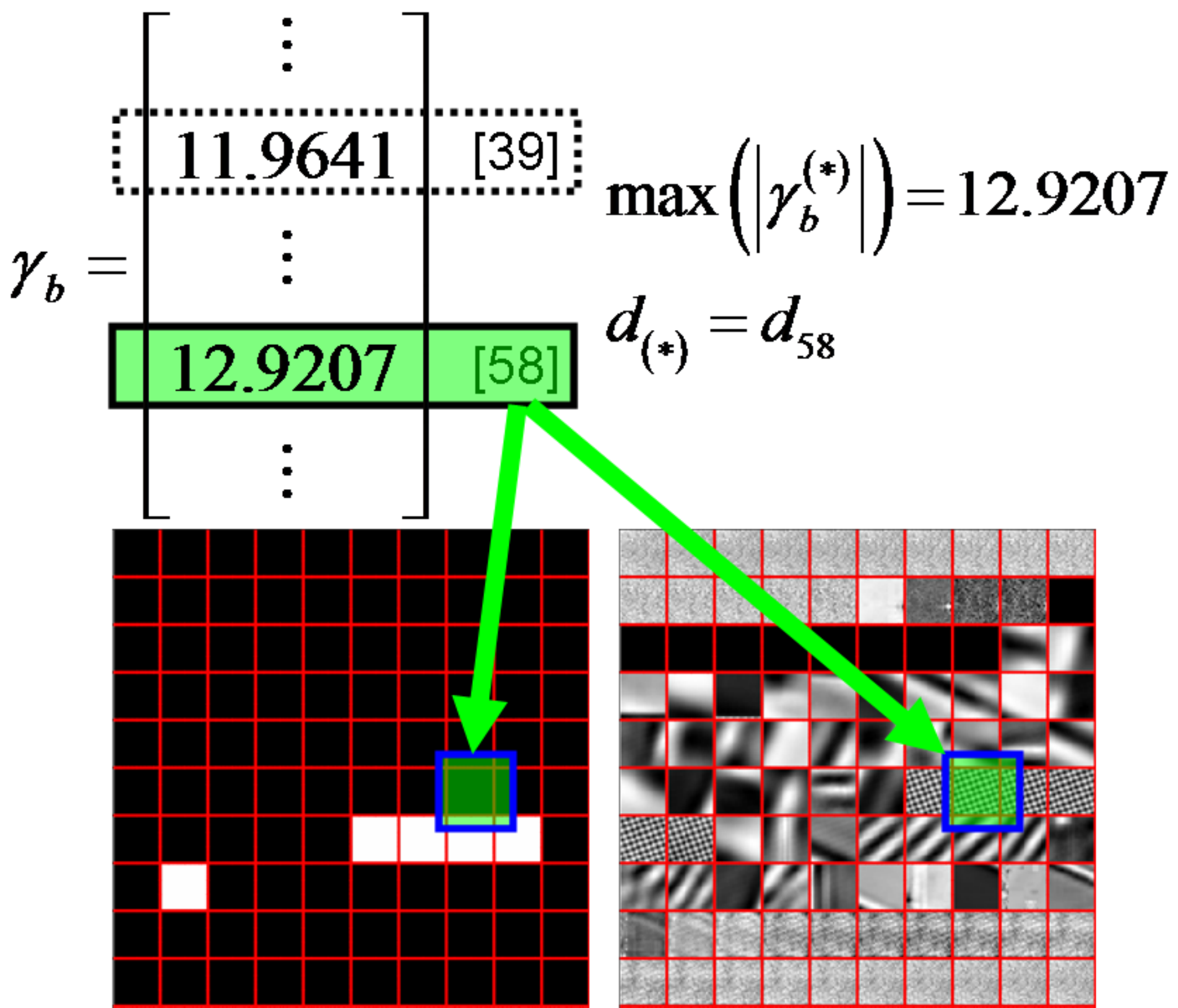}
        \label{show_sparse_coefficient_seg:back_sparse_vec}
    }
    \caption{The illustration of the details during the automated sparse coefficient-based ROI segmentation procedure: (a) the two types of image blocks in latent print image - the foreground block (marked by pink square) and the background block (marked by green square); (b) the marked foreground block; (c) seeking for the specific atom corresponding to the $max({\left| {\gamma_f^{\left(  *  \right)}} \right|})$ for (b) (the found atom is identified as the ``ridge-valley" atom in preceding phase); (d) the marked background block; and (e) seeking for the specific atom corresponding to the $max({\left| {\gamma_b^{\left(  *  \right)}} \right|})$ for (d) (the found atom is labeled as the ``non-ridge-valley" atom in preceding phase).}
    \label{show_sparse_coefficient_seg}
\end{figure}
%------------------- show foreground / background latent block -------------------

\begin{figure}
    \centering
    \subfigure[]
    {
        \includegraphics[height=1.5in]{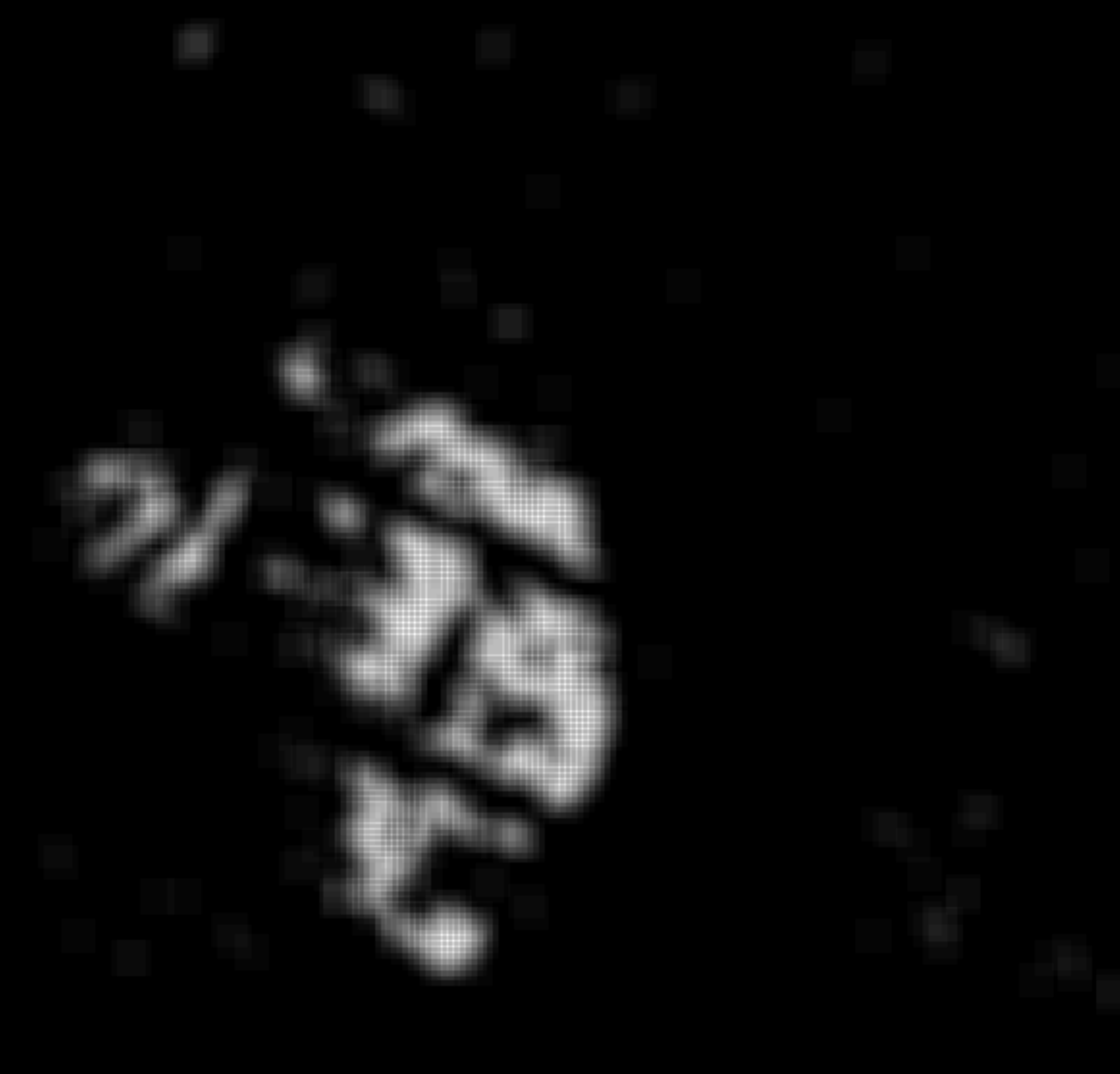}
        \label{show_sparse_coefficient_ROI:M}
    }
    \subfigure[]
    {
        \includegraphics[height=1.5in]{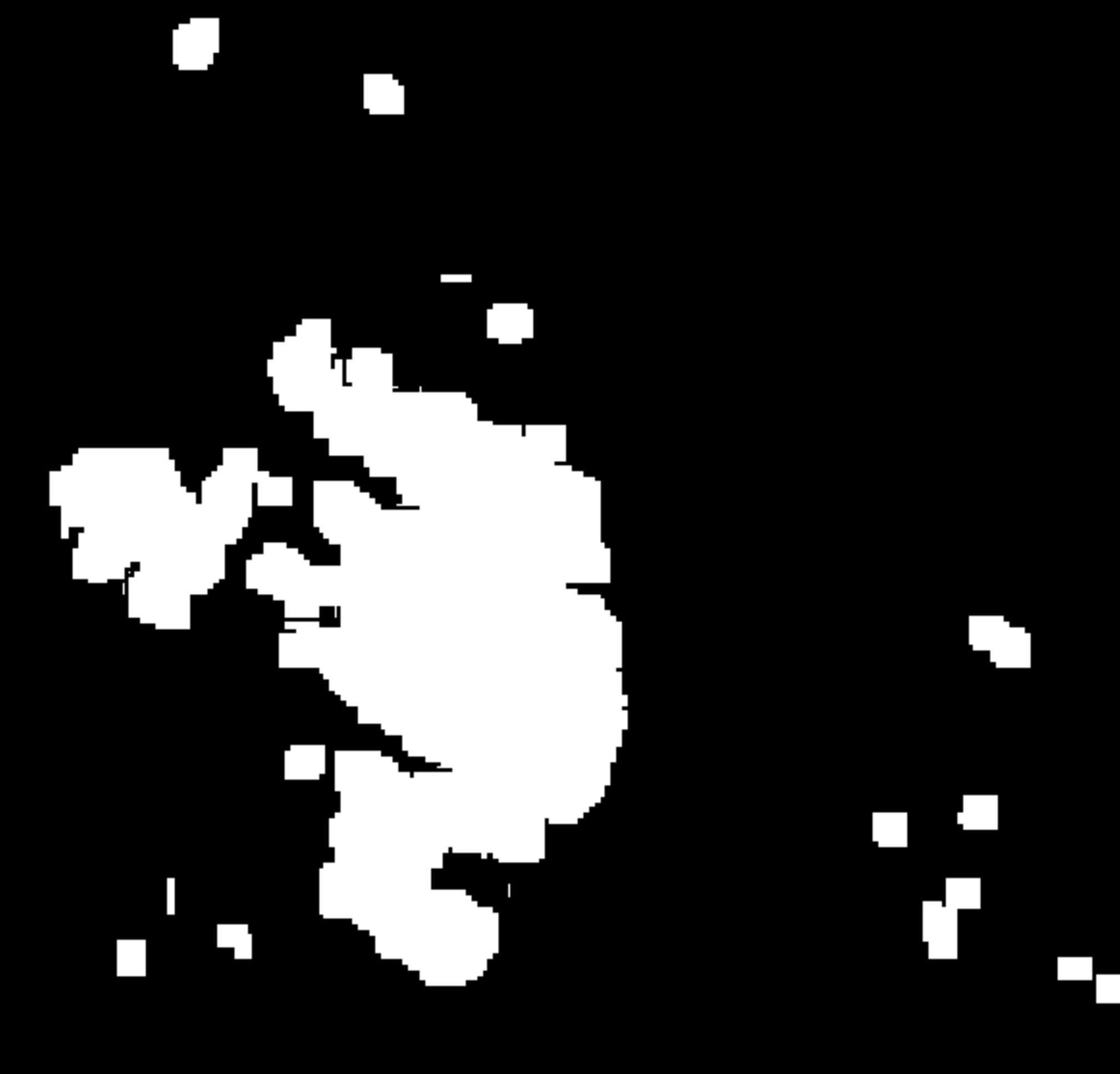}
        \label{show_sparse_coefficient_ROI:OTSU}
    }
    \subfigure[]
    {
        \includegraphics[height=1.5in]{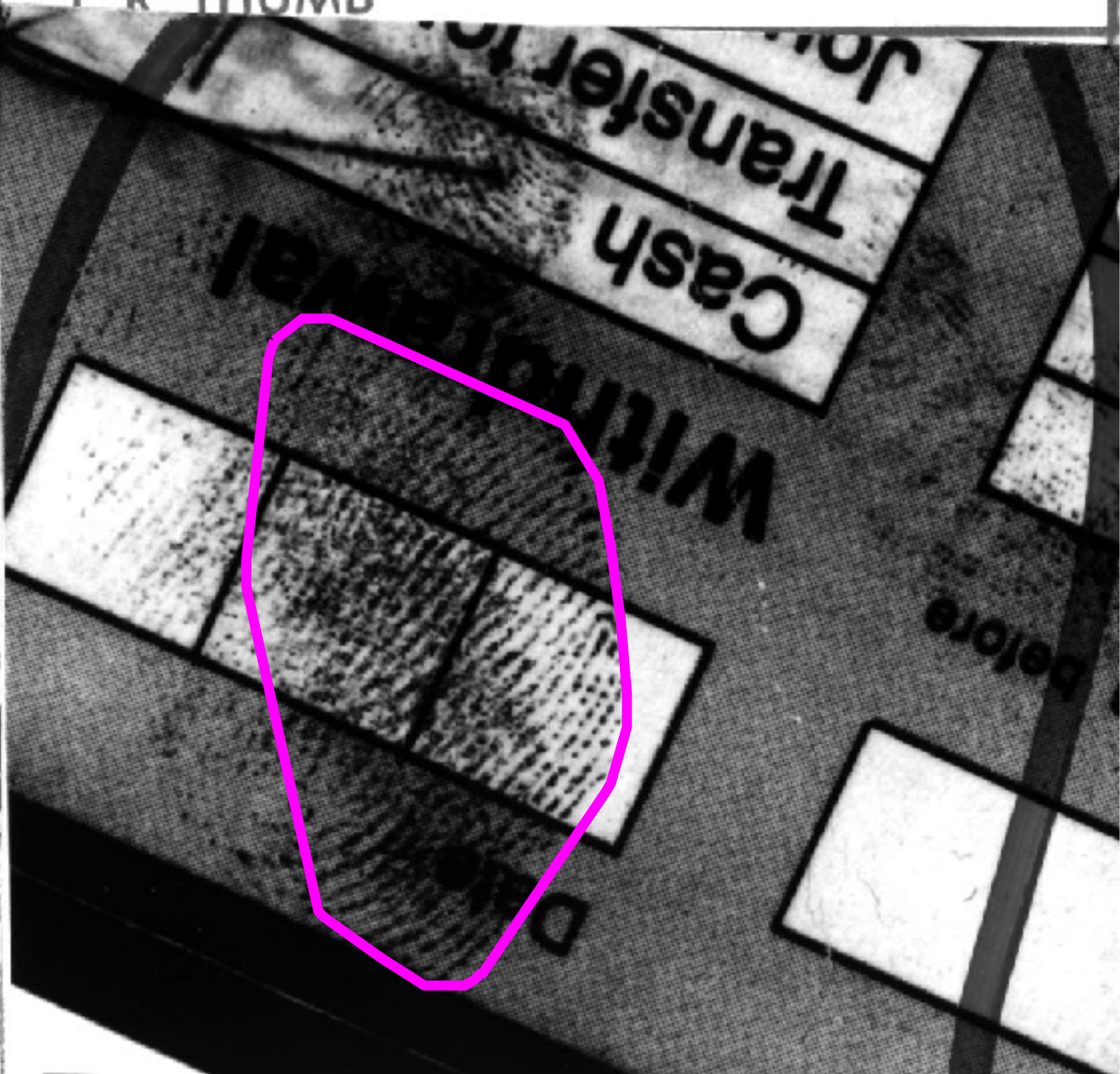}
        \label{show_sparse_coefficient_ROI:ROI}
    }
    \caption{The example of the ROI obtained by using the proposed sparse coefficient-based ROI segmentation procedure: (a) the image $M$ obtained after all the signal vectors have been processed; (b) the binary image obtained by using the normalization and Otsu's adaptive threshold-based segmentation method; and (c) the polygonal ROI obtained via the mathematical morphology operation and the maximal-area convex polygon extraction (marked by pink convex polygon over the original latent print image U224 in NIST SD27).}
    \label{show_sparse_coefficient_ROI}
\end{figure}

\begin{itemize}

  \item {\it Step 1:} Initialize a vacant image $M$ (all the pixel intensities are zero) with the same size as the original latent fingerprint image, then divide the latent image into the overlapping blocks (block size $w \times w$, as shown in Figure \ref{show_sparse_coefficient_seg:fore_patch} and Figure \ref{show_sparse_coefficient_seg:back_patch}) and rearrange the obtained image blocks into the column vectors (here the training set $S = \left\{ {{s_i}\left| {i = 1,2,...,N} \right.} \right\}$ for dictionary learning is directly used, where $s_i$ is obtained after the vectorization for the image block $p_i$);

  \item {\it Step 2:} Given a signal vector $s_i$, orthogonal matching pursuit (OMP) \cite{Tropp04} is applied to the given vector $s_i$ for selecting few atoms to sparsely approximate $s_i$, then the sparse coefficient vector $\gamma_i$ is obtained;

  \item {\it Step 3:} Find the highest nonzero absolute value $\left| {\gamma _i^{\left( * \right)}} \right|$ in sparse coefficient vector $\gamma_i$ and the corresponding atom ${d_{\left(  *  \right)}}$ in dictionary $D$ (shown in Figure \ref{show_sparse_coefficient_seg:fore_sparse_vec} and Figure \ref{show_sparse_coefficient_seg:back_sparse_vec});

  \item {\it Step 4:} Check whether the found atom ${d_{\left(  *  \right)}}$ is the identified ``ridge-valley" atom (shown in Figure \ref{show_sparse_coefficient_seg:fore_sparse_vec} and Figure \ref{show_sparse_coefficient_seg:back_sparse_vec}). Further, the pixel intensities inside the block of $M$ whose size and block-wise coordinate are the same as the currently processed block $p_i$ in latent image, are tuned according to Equation (\ref{M_map}).

      \begin{equation}
      \label{M_map}
      M_{p_i}(x,y) =
      \begin{cases}
      M_{p_i}(x,y) + 1, \ d_{(*)} is ``ridge-valley" atom \\
      M_{p_i}(x,y), \ d_{(*)} is \ not ``ridge-valley" atom
      \end{cases}
      \end{equation}

      where ${M_{{p_i}}}\left( {x,y} \right)$ denotes the pixel intensities inside the block of $M$, whose size and block-wise coordinate are the same as the currently processed block $p_i$ in latent image.

%      \newcounter{mytempeqncnt2}
%      \begin{figure*}[!t]
%      \normalsize
%      \setcounter{mytempeqncnt2}{\value{equation}}
%      \setcounter{equation}{12}
%      \begin{equation}
%      \label{M_map}
%      M_{p_i}(x,y) =
%      \begin{cases}
%      M_{p_i}(x,y) + 1, & \text {\it if} \ \ d_{(*)} \ \ {\it is \ ``ridge-valley" \ atom} \\
%      M_{p_i}(x,y), & \text {\it otherwise}
%      \end{cases}
%      \end{equation}
%      \setcounter{equation}{\value{mytempeqncnt2}}
%      \hrulefill
%      \vspace*{4pt}
%      \end{figure*}

  \item {\it Step 5:} Check whether all the signal vectors have been processed. If not, go back to {\it Step 2}. Otherwise, terminate the calculation of sparse coefficient for signal vector and output the image $M$ (shown in Figure \ref{show_sparse_coefficient_ROI:M});

  \item {\it Step 6:} All the pixel intensities in image $M$ need to be normalized to the range $[0, 1]$, then the normalized image is binarized by Otsu's adaptive threshold-based segmentation method \cite{Otsu1979} (shown in Figure \ref{show_sparse_coefficient_ROI:OTSU});

  \item {\it Step 7:} A set of mathematical morphology operators such as close, open, hole-filling, and small block removal are applied to the obtained binary image, finally the maximal-area convex polygon containing the foreground regions is achieved as the ROI (shown in Figure \ref{show_sparse_coefficient_ROI:ROI}).

\end{itemize}

%------------------- show ROI -------------------
%------------------- show ROI -------------------

\subsection{Genetic Algorithm-Based Latent Matching Unit}

\subsubsection{ROI-Based Minutiae Extraction}

The minutiae sets for latent fingerprint images are extracted from the ROI obtained in segmentation module. As demonstrated in \cite{Paulino11} \cite{Paulino13} \cite{Jain08} and \cite{Jain11}, the reliability of minutiae plays a crucial role in the performance of latent matcher. Obviously, the involvement of spurious minutiae deteriorates the matching performance. That is, the more spurious involved, the poorer the matcher behaves. However, how to effectively reduce or even avoid the spurious minutiae in fully-automatic mode becomes a significant issue. As the outcome of segmentation scheme, the resultant ROI is used for automated minutiae extraction which can not only effectively preserve the genuine minutiae but also significantly reduce the false ones. That is, the genuine minutiae are most likely to be detected within ROI while the spurious ones caused by the structured noise in background are effectively eliminated by ROI mask. Therefore, the minutiae set obtained from ROI in latent image is supposed to be reliable.

\subsubsection{Problem Formulation for Minutiae Set-Based Fingerprint Matching}

The minutiae set-based fingerprint matching can be mathematically formulated as follows: let $C = \left\{ {{c_i}\left| {i = 1,2,...,m} \right.} \right\}$ and $L = \left\{ {{l_j}\left| {j = 1,2,...,n} \right.} \right\}$ be the two point set in ${R^2}$ space ($C$ stands for the minutiae set extracted from the currently compared print and $L$ denotes the minutiae set extracted from the ROI in query latent print respectively). Both sets determine whether there is a specific affine transformation $T = \left( {\theta ,s,{t_x},{t_y}} \right)$ ($\theta$ denotes the rotation angle, $s$ represents the scaling factor, and $t_x$ and $t_y$ are the offsets along $x-$ and $y-$ axis respectively) that maps set $C$ onto or close to set $L$ to indicate a correspondence. Therefore, seeking for the affine transformation as well as the correspondences (an exact one-to-one correspondence, or an approximate correspondence) between both point sets $C$ and $L$ are coupled point matching problem. Here let $\left( {{c_i},{l_j}} \right)$ be one of the corresponding pairs under $T$, and denote ${c_i} = {( {{x_{{c_i}}},{y_{{c_i}}},{o_{{c_i}}},p{t_{{c_i}}}} )^T}$ and ${l_j} = {( {{x_{{l_j}}},{y_{{l_j}}},{o_{{l_j}}},p{t_{{l_j}}}} )^T}$. $(x_{c_i}, y_{c_i})$ and $(x_{l_j}, y_{l_j})$ are corresponding coordinates, $o_{c_i}$, $o_{l_j}$, $pt_{c_i}$ and $pt_{l_j}$ denote the point orientation and type of $c_i$ and $l_j$ respectively. The formula of affine transformation is denoted as follows

\begin{equation}
\left( {{x_{{l_j}}},{y_{{l_j}}}} \right) = T\left[ {\left( {{x_{{c_i}}},{y_{{c_i}}}} \right)} \right]
\end{equation}

\begin{equation}
\left[ {\begin{array}{*{20}{c}}
{{x_{{l_j}}}}\\
{{y_{{l_j}}}}
\end{array}} \right] = s \cdot \left[ {\begin{array}{*{20}{c}}
{\cos \theta }&{ - \sin \theta }\\
{\sin \theta }&{\cos \theta }
\end{array}} \right] \cdot \left[ {\begin{array}{*{20}{c}}
{{x_{{c_i}}}}\\
{{y_{{c_i}}}}
\end{array}} \right] + \left[ {\begin{array}{*{20}{c}}
{{t_x}}\\
{{t_y}}
\end{array}} \right]
\label{affine_transformation}
\end{equation}

Equation (\ref{affine_transformation}) indicates that the coupled point matching problem can be regarded as the parameter tuning problem. That is, with the optimal selection for the parameter, the corresponding affine transformation is generated. Given the optimal affine transformation, set $C$ are mapped onto set $L$. After transformation, the most points in set $C$ being close to the points in set $L$ are regarded as matched points.

\subsubsection{Minutiae Set-Based Matching Problem Solved by GA}

Seeking for the appropriate affine transformation parameters is important for the alignment between latent minutiae set and the compared print minutiae set. Such suitable parameters ensure the largest overlap of global topology between two point sets. Considering that the affine transformation parameters need to be elaborately tuned, thus the parameter optimization technique is necessary. In contrast to the deterministic optimization algorithm like greedy searching method proposed in \cite{Jain08} and \cite{Jain11}, the evolutionary optimization approaches have the advantage of computational efficiency and the capability to effectively avoid the local optimum. As a consequence, GA, one of the typical evolutionary optimization approaches, is applied to solve the minutiae set-based matching problem.

%GA was proposed by John Holland in 1975, which is an adaptive heuristic search algorithm according to the evolutionary idea of natural selection \cite{Agoston03}. Due to its capability of heuristic searching, GA is utilized to solve global optimization problems.
To be specific, GA begins with the randomly initialized chromosomes which represent the solution of problem. In subsequent iteration, the updated chromosomes are obtained by using various genetic operators. According to the fitness function, the previous chromosomes are substituted by the updated ones when the updated ones are judged to be fitter individuals than the previous ones. With the continuous iteration, the chromosomes are motivated to evolve to the fittest individuals until the termination of algorithm. Because GA is not easy to trap into local optimal and its searching manner is potentially parallel, it has been broadly utilized to solve the point set matching problem \cite{Li03} \cite{Zhang03} and \cite{Xu2013}.

In order to use GA, the affine transformation parameters, namely rotation angle $\theta$, scaling factor $s$, and translation offsets $t_x$ and $t_y$ are coded as chromosome. The chromosome vector is denoted as follow

\begin{equation}
Chrom = {\left( {\theta ,s,{t_x},{t_y}} \right)^T}
\end{equation}

where each parameter value is a random real number and is restricted in an appropriate range. Compared with the binary coding, the real number-based coding has the following advantages: effectively avert the hamming cliff and avoid the decimal digits assignment. Each encoded chromosome corresponds to one parameter set of affine transformation to align the point set. During the iterations of GA, the chromosomes are constrained in the same range as they are randomly initialized.

To motivate the evolution of GA, the fitness of each chromosome, needs to be evaluated. In order to define the fitness function, coordinate distance $e_d$ and point orientation difference $e_o$ need to be calculated in advance

\begin{equation}
{e_d} = \sqrt {{{\left[ {T\left( {{x_{{c_i}}}} \right) - {x_{{l_j}}}} \right]}^2} + {{\left[ {T\left( {{y_{{c_i}}}} \right) - {y_{{l_j}}}} \right]}^2}}
\label{distance_measure}
\end{equation}

\begin{equation}
{e_o} = \left| {T\left( {{o_{{c_i}}}} \right) - {o_{{l_j}}}} \right|
\label{orientation_measure}
\end{equation}

where ${T\left( {{o_{{c_i}}}} \right)}$ is point $c_i$'s orientation after rotation caused by transformation $T$. Given the resultant $e_d$ and $e_o$, the fitness function is defined as follow

\begin{equation}
{num}^t = \begin{cases}
{num}^{t-1} + 1, & \text{\it if}\ {e_d} \le {\delta _d}, {e_o} \le {\delta _o}, p{t_{{c_i}}} = p{t_{{l_j}}} \\
{num}^{t-1}, & \text{\it otherwise}
\end{cases}
\label{fitness_function}
\end{equation}

where ${num}^t$ and ${num}^{t-1}$ are the numbers of matched point pairs in $t^{th}$ and ${(t-1)}^{th}$ iteration respectively. Therein, the number of paired points is directly assigned as the fitness function value. ${\delta _d}$ and ${\delta _o}$ are the tolerances for $e_d$ and $e_o$ respectively. Equation (\ref{fitness_function}) indicates that for every iteration all the matched point pairs need to be found and counted when the following three criterions are simultaneously satisfied: (i) ${{e_d} \le {\delta _d}}$; (ii) ${{e_o} \le {\delta _o}}$; and (iii) ${p{t_{{c_i}}} = p{t_{{l_j}}}}$. Therefore, a chromosome producing larger $num$ (higher fitness) is considered to be superior to the other chromosomes with smaller $num$ (lower fitness). As such, the fittest chromosomes can be determined according to the fitness values of all chromosomes. In addition, Equations (\ref{distance_measure}-\ref{fitness_function}) not only consider the global topology of point set but also involve the point property such as point orientation and type during the iterative evolution of GA.

To boost the evolution of GA, the GA operators such as selection, crossover and mutation are also important. To be more specific, the selection operator is used to select the chromosomes with higher fitness, which preserve and inherit the information of these fitter chromosomes into next generation. Therein, the widely used Roulette Wheel selection scheme is applied and the selection probability of each chromosome is proportional to its own fitness value. Besides, the creation of new offsprings to enhance the diversity of chromosomes is necessary because only the inheritance of fitter chromosomes in iterative evolution is not sufficient. Therefore, crossover operator produces new chromosomes through combining partial segments of two parent chromosomes. Therein, the multi-point crossover operator is adopted. Moreover, in order to further enhance chromosomes diversity, the uniform mutation operator is performed to randomly shift the value of chromosome vector with a small probability.

\begin{figure}
    \centering
%    \subfigure[]
%    {
%        \includegraphics[height=1.5in]{show_GA_matching_tenprint_veri.pdf}
%        \label{show_GA_genuine_matching_result:tenprint_veri}
%    }
%    \subfigure[]
%    {
%        \includegraphics[height=1.5in]{show_GA_matching_latent_veri.pdf}
%        \label{show_GA_genuine_matching_result:latent_veri}
%    }
    \subfigure[]
    {
        \includegraphics[height=2in]{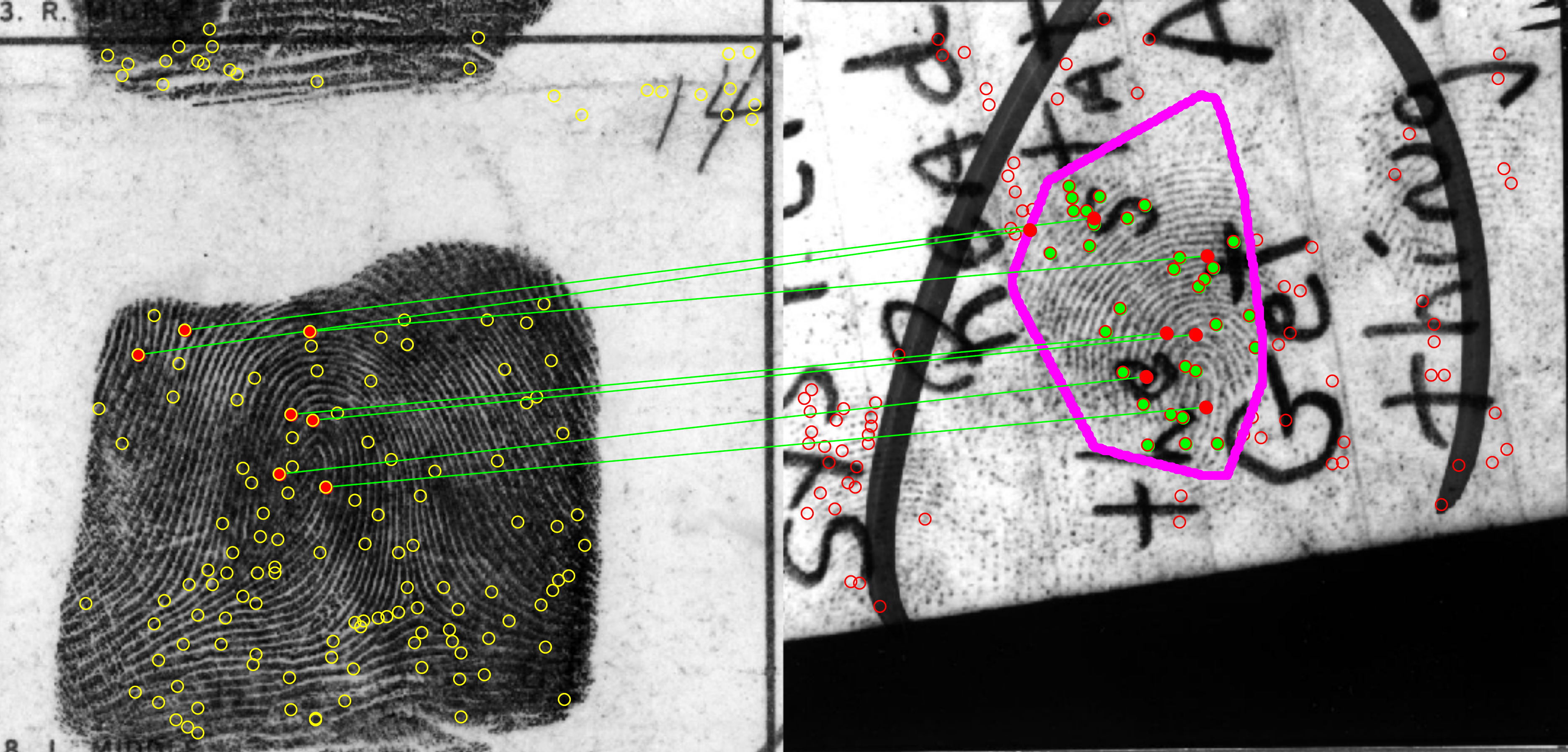}
        \label{show_GA_genuine_matching_result:result}
    }
    \subfigure[]
    {
        \includegraphics[height=2in]{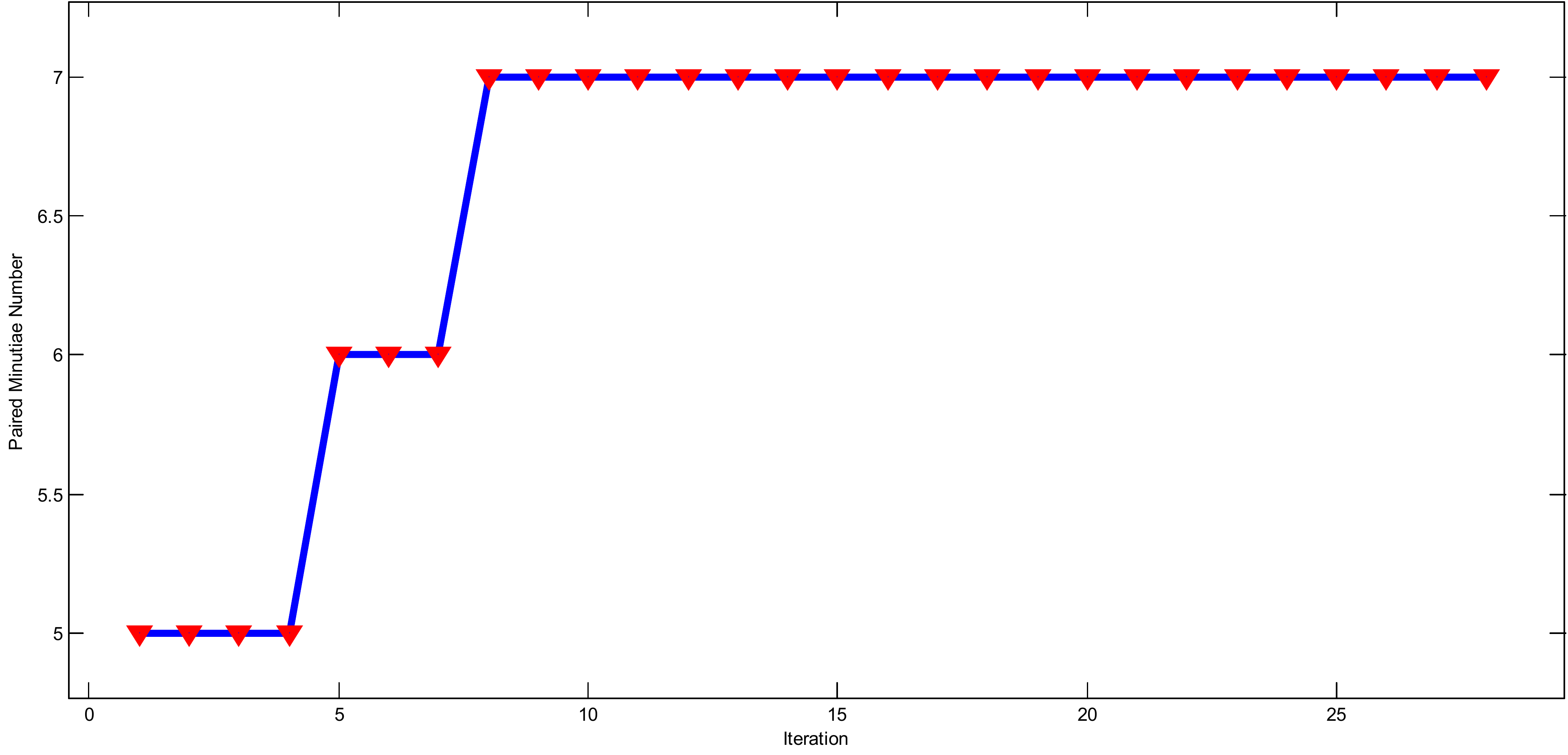}
        \label{show_GA_genuine_matching_result:iteration}
    }
    \caption{The example of successful matching for genuine ``rolled-latent" pair by proposed GA-based matching unit. In this case, seven minutiae pairs are found. The paired minutiae between the rolled and latent print are connected by green lines and marked by red solid points respectively. (a) the rolled print image (left) corresponds to G009 in NIST SD27 (right); the automatically extracted minutiae in rolled print image are marked by yellow hollow points and the ones in latent print image are marked by red hollow points; the ROI is highlighted by pink convex polygon and the automatically extracted minutiae within the ROI are marked by green solid points; and (b) the augment of the paired minutiae number with the iteration of GA.}
%The example of successful matching for genuine ``rolled-latent" pair by proposed GA-based matching unit. In this case, seven minutiae pairs are found. The paired minutiae between the rolled and latent print are connected by green lines and marked by red solid points respectively. (a) the rolled print image (left) corresponding to G009 in NIST SD27 (right); the automatically extracted minutiae in and the automatically extracted minutiae (marked by yellow hollow points); (b) the latent print image G009 and the ROI-based automatically extracted minutiae (the ROI is highlighted by pink convex polygon and the automatically extracted minutiae within the ROI are marked by green solid points; the red hollow points represent the automatically extracted minutiae outside the ROI); (c) the matching result for the genuine ``rolled-latent" pair (seven minutiae pairs are found; the paired minutiae between the rolled and latent print are connected by green lines and marked by red solid points respectively); and (d) the augment of the paired minutiae number with the iteration of GA.
    \label{show_GA_genuine_matching_result}
\end{figure}

In this study, the GA-based minutiae set matching algorithm is summarized as follows:

\begin{itemize}

  \item {\it Step 1:} Extract the minutiae set $C$ from the currently compared print and the minutiae set $L$ from the ROI in query latent print (shown in Figure \ref{show_GA_genuine_matching_result:result}), then set population size of chromosomes $S_{chrom}$, crossover probability $p_c$, mutation probability $p_m$, the maximal iterations $g_{max}$, point coordinate distance tolerance $\delta_d$, point orientation difference tolerance $\delta_o$ and the value range for chromosome vector;

  \item {\it Step 2:} Randomly generate chromosomes within the value range as initial generation;

  \item {\it Step 3:} Compute fitness values for all chromosomes in current generation and then select the fitter ones by selection operator;

  \item {\it Step 4:} Apply crossover operator and mutation operator to the fitter chromosomes, then create the new chromosomes for next generation;

  \item {\it Step 5:} Check whether the maximal iterations is reached, or check whether the highest fitness values are maintained the same during several iterations. If not reached or maintained, go back to {\it Step 3}. Otherwise terminate GA and output the fittest chromosome as the estimated optimal parameter for affine transformation;

  \item {\it Step 6:} Align minutiae set $C$ versus minutiae set $L$ based on the obtained transformation parameter, then seek for the corresponding minutiae point pairs according to the criterions in Equation (\ref{fitness_function}) (shown in Figure \ref{show_GA_genuine_matching_result:result});

  \item {\it Step 7:} Output the number of paired minutiae between $C$ and $L$ as the matching score when it has converged as shown in Figure \ref{show_GA_genuine_matching_result:iteration}.

  %The matching score is expected to be as high as possible for the genuine ``rolled-latent" pairs (a successful matching for genuine ``rolled-latent" pair is shown in Figure \ref{show_GA_genuine_matching_result}) while to be as low as possible for the imposter ``rolled-latent" pairs (an unsuccessful matching for imposter ``rolled-latent" pair is illustrated in Figure \ref{show_GA_imposter_matching_result}).

\end{itemize}

%------------------- show GA-based genuine matching result -------------------
%------------------- show GA-based genuine matching result -------------------

%------------------- show GA-based imposter matching result -------------------
%\begin{figure}
%\centerline{\includegraphics[height=1.5in]{show_GA_matching_imposter_result.pdf}}
%\caption{The example of unsuccessful matching for imposter ``rolled-latent" pair by proposed GA-based matching unit: the rolled print corresponding to G001 in NIST SD27 is matched against the latent print G009; no one minutiae pair can be found after the termination of GA iteration.}
%\label{show_GA_imposter_matching_result}
%\end{figure}
%------------------- show GA-based imposter matching result -------------------

\section{Experimental Results}

In this section, the introduced latent fingerprint matcher has been evaluated by the following two experiments: (i) the minutiae extraction based on the ROI; and (ii) the latent fingerprint matching.

\subsection{Data Preparation}

All the experiments are conducted on the latent fingerprint database NIST SD27 which is available in the public domain. Such database includes $258$ latent print images and their corresponding rolled print images, where these $258$ latent images are grouped by the latent examiners into the following three categories: ``Good", ``Bad" and ``Ugly". The numbers of latent images involved in ``Good", ``Bad" and ``Ugly" categories are $88$, $85$ and $85$ respectively.

\subsection{Experiment 1: ROI-Based Minutiae Extraction}

In this experiment, the reliability of automated minutiae extraction based on the ROI in latent print is evaluated. All the minutiae within the ROI are automatically extracted by VeriFinger SDK. Given query latent print image, the following three scenarios are compared.

\begin{itemize}

  \item {\bf Minutiae Extraction Scenario 1 - Whole Image-Based Automated Minutiae Extraction:} instead of the segmentation for the foreground, the whole latent image is directly used for the automated minutiae extraction.

  \item {\bf Minutiae Extraction Scenario 2 - Proposed ROI Segmentation Module-Based Automated Minutiae Extraction:} the ROI is obtained by the proposed dictionary learning-based ROI segmentation module and then used for the automated minutiae extraction. The parameters used in the proposed ROI segmentation module are empirically tuned as follows: the patch size for training sample selection $w = 32$, the atom number $N_a = 100$, the sparsity parameter $K = 2$, and the cross-correlation threshold to judge the pattern similarity $T{h_{xcorr}} = 0.6$.

  \item {\bf Minutiae Extraction Scenario 3 - Cao's ROI Segmentation Approach-Based Automated Minutiae Extraction:} the ROI is obtained by \cite{Cao14} and then used for the automated minutiae extraction. The parameters used in this method are tuned as the suggested ones.

\end{itemize}

In order to assess the performance of the ROI-based automated minutiae extraction, the manually marked minutiae provided by NIST SD27 for query latent fingerprint are used as the ground-truth to label the genuine minutiae. To be more specific, ${MS}_1$ is the ground-truth minutiae set (provided by NIST SD27 and marked by the FBI latent fingerprint examiners); ${MS}_2$ is the minutiae set automatically extracted from the whole latent image by VeriFinger SDK; and ${MS}_3$ is the minutiae set automatically extracted from the ROI by VeriFinger SDK. Accordingly, two metrics such as genuine minutiae preservation rate (GMPR) and false minutiae acceptance rate (FMAR) are defined as follows

%\begin{figure}
%\centerline{\includegraphics[scale=0.38]{GPR_FAR_set_relation_v2.pdf}}
%\caption{The relationships among the Ground-Truth minutiae set (${MS}_1$), the minutiae set automatically extracted from the whole latent image (${MS}_2$), and the minutiae set automatically extracted from the ROI after segmentation procedure (${MS}_3$).}
%\label{GPR_FAR_set_relation}
%\end{figure}

\begin{equation}
GMPR = \frac{{\# \left\{ {M{S_1} \cap M{S_3}} \right\}}}{{\# \left\{ {M{S_1} \cap M{S_2}} \right\}}}
\label{gmpr_equ}
\end{equation}

\begin{equation}
FMAR = \frac{{\# \left\{ {M{S_3} - M{S_1} \cap M{S_3}} \right\}}}{{\# \left\{ {M{S_2} - M{S_1} \cap M{S_2}} \right\}}}
\label{fmar_equ}
\end{equation}

where GMPR is defined as the percentage of the minutiae belonging to the genuine minutiae set which are also correctly extracted by computer program. FMAR is defined as the percentage of the minutiae belonging to the structured noise in latent image which are wrongly detected as the genuine ones by computer program. $\# \left\{ (*) \right\}$ denotes the number of minutiae included in set $(*)$. Under the condition of fully automated minutiae extraction, the effect of ROI is evaluated depending on the benchmark where no ROI mask is adopted but the whole query latent image is used for automated minutiae extraction. Therefore, in Equations (\ref{gmpr_equ}) and (\ref{fmar_equ}), not the manually marked genuine minutiae $\# \left\{ {M{S_1}} \right\}$ but the automatically extracted genuine minutiae $\# \left\{ {M{S_1} \cap M{S_2}} \right\}$ and $\# \left\{ {M{S_2} - M{S_1} \cap M{S_2}} \right\}$ are used as the baseline. For Scenario 1, ${M{S_1} - M{S_1} \cap M{S_2}}$ stands for the missing genuine minutiae, even the entire latent image is imported into the automated minutiae extractor; $M{S_1} \cap M{S_2}$ represents the genuine minutiae which are correctly detected by the computer program; and ${M{S_2} - M{S_1} \cap M{S_2}}$ are the spurious minutiae falsely extracted by the computer program. For Scenario 2 and 3, ${M{S_1} - M{S_1} \cap M{S_3}}$ denotes the missing genuine minutiae after the adoption of ROI; $M{S_1} \cap M{S_3}$ stands for the genuine minutiae correctly detected by the computer program inside ROI domain; and ${M{S_3} - M{S_1} \cap M{S_3}}$ represents the false minutiae wrongly extracted by the computer program within ROI.

The minutiae extracted by the automated computer program could not be exactly the genuine ones and unavoidably contain fakes. That is, image-based minutiae detection and extraction is primarily depending on the locally salient image structures such as ridge ending or ridge bifurcation in image domain. For the latent image, such typical structures are not distinct or even lost due to the low clarity of ``ridge-valley" pattern in fingerprint region, and consequently lead to the loss of genuine minutiae; some other image components in background also have the similar ridge ending or bifurcation structures, and accordingly yield the spurious minutiae. Therefore, the missing detection for the genuine minutiae and the false detection for the imposter ones by the automated computer program (e.g. VeriFinger SDK) are usually unavoidable. As a consequence, for Scenario 1, the sets ${M{S_1} - M{S_1} \cap M{S_2}}$ and ${M{S_2} - M{S_1} \cap M{S_2}}$ are not empty. Similarly, for Scenario 2 and 3, ${M{S_1} - M{S_1} \cap M{S_3}}$ and ${M{S_3} - M{S_1} \cap M{S_3}}$ are not empty either.

%------------------- show minutiae extraction result -------------------
\begin{figure}
    \centering
    \subfigure[]
    {
        \includegraphics[height=2in]{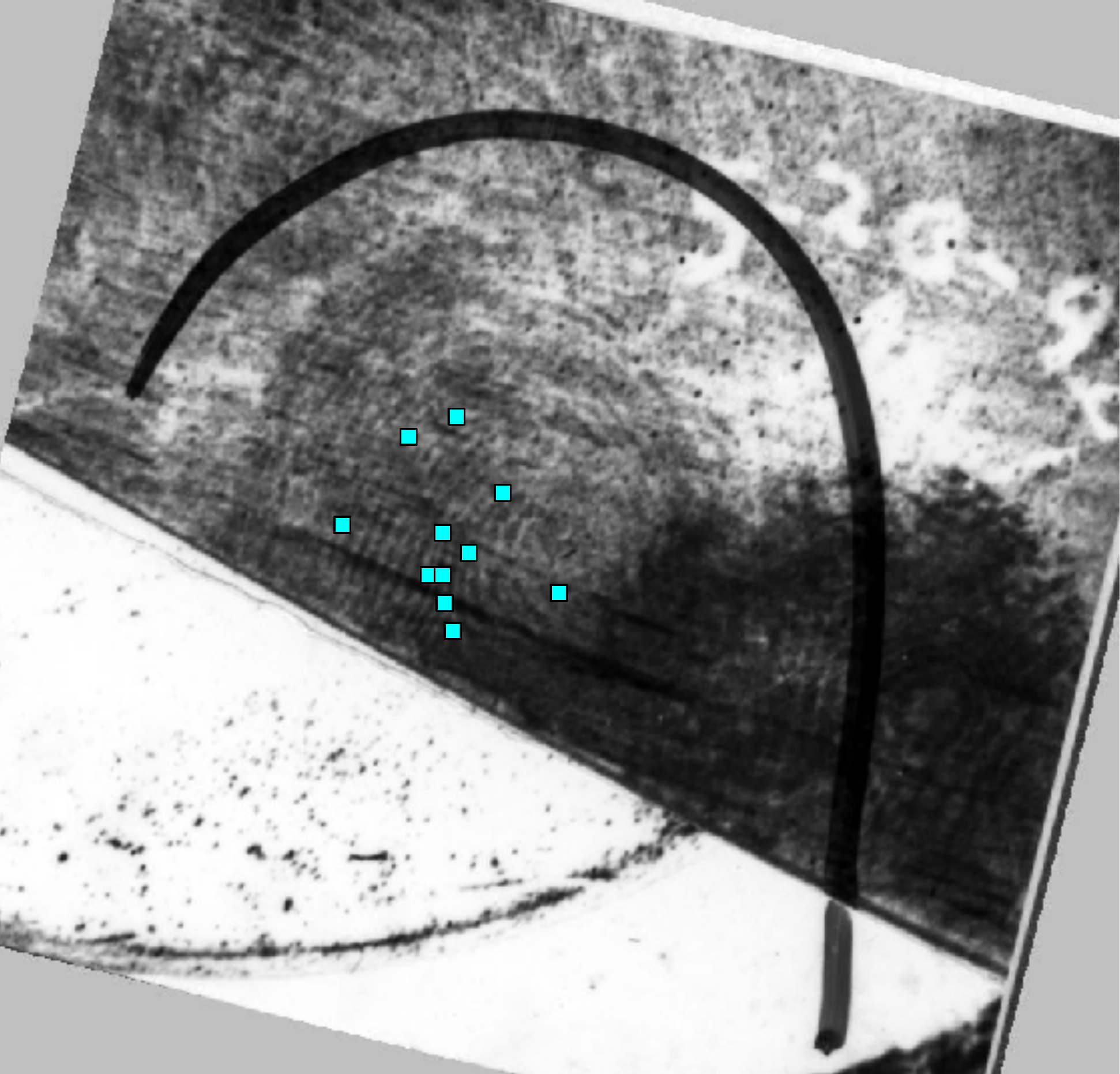}
        \label{show_minutiae_extraction:s0}
    }
    \subfigure[]
    {
        \includegraphics[height=2in]{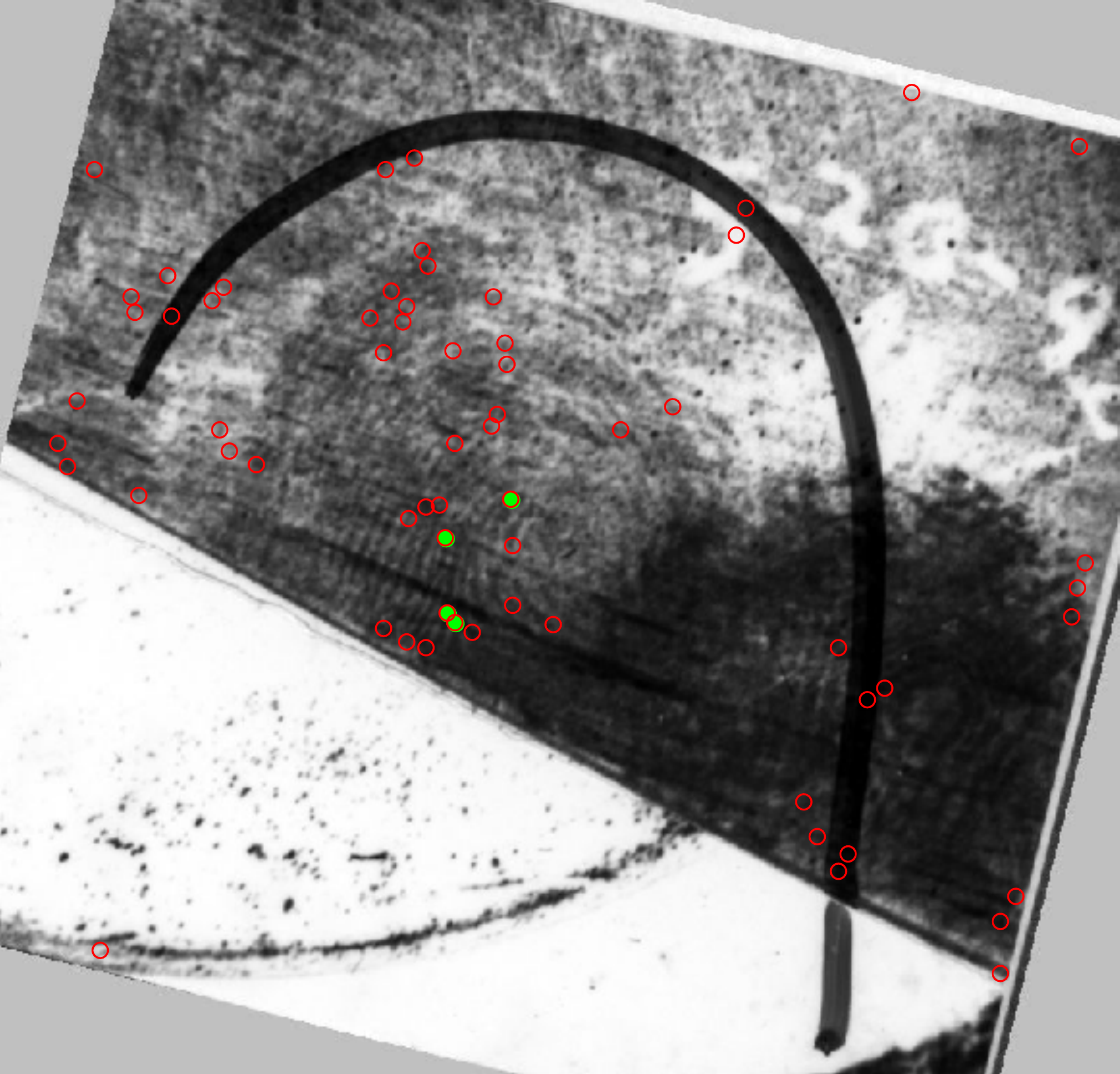}
        \label{show_minutiae_extraction:s1}
    }
    \\
    \subfigure[]
    {
        \includegraphics[height=2in]{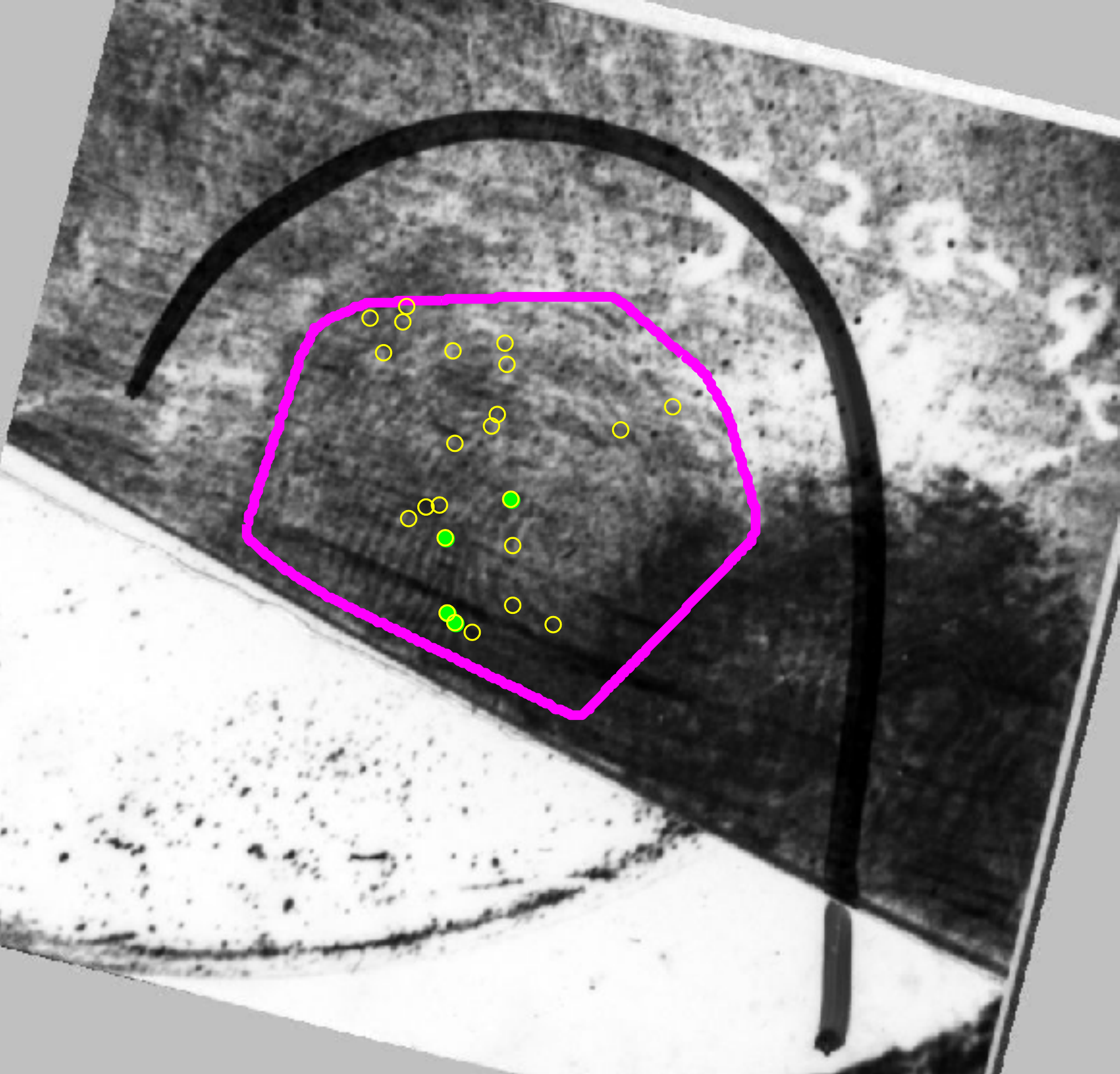}
        \label{show_minutiae_extraction:s2}
    }
    \subfigure[]
    {
        \includegraphics[height=2in]{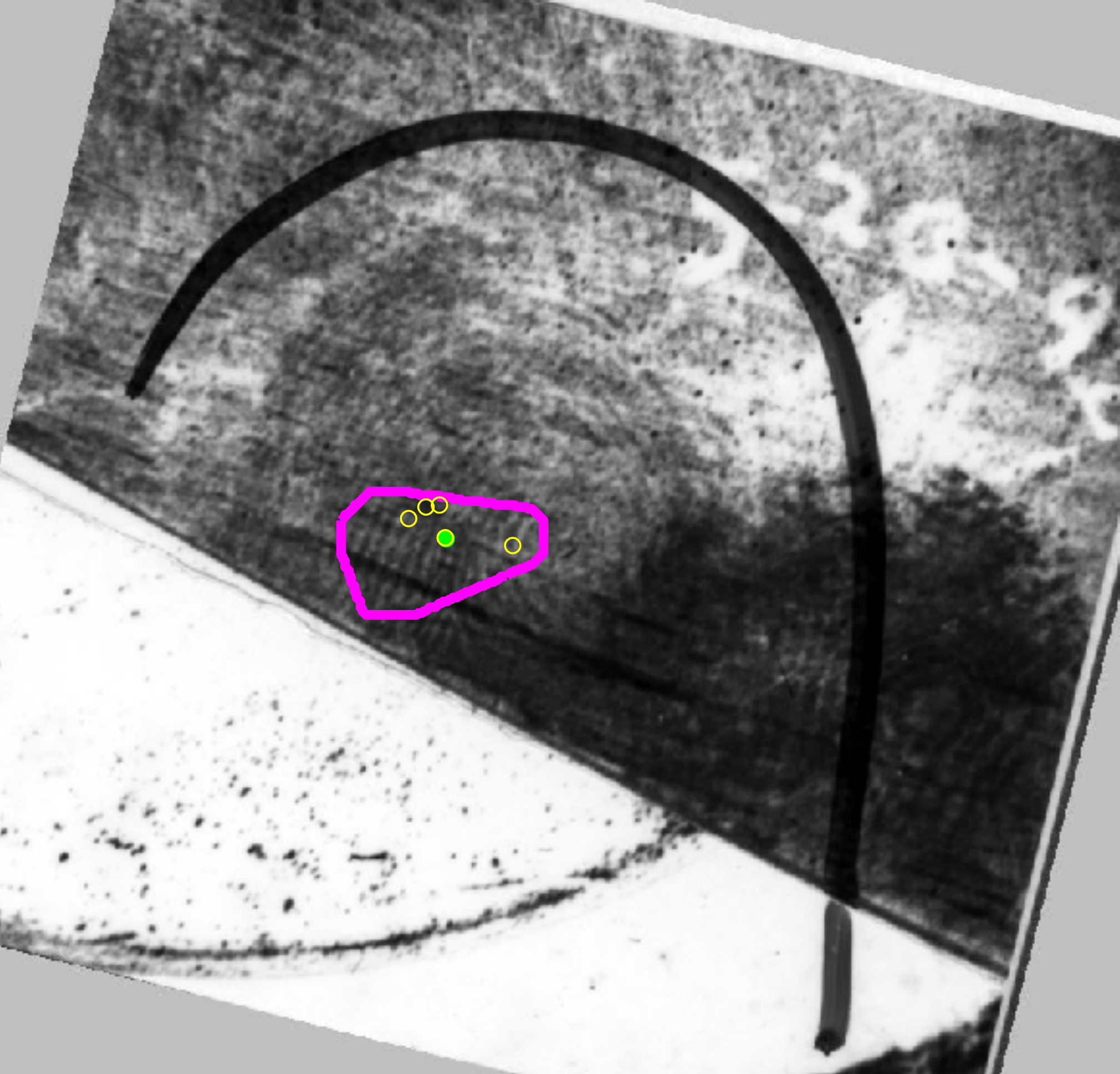}
        \label{show_minutiae_extraction:s3}
    }
    \caption{The example of three scenarios on U267 in NIST SD27: (a) the ground-truth minutiae set provided by NIST SD27; (b) Scenario 1 - four genuine minutiae can be correctly extracted, $GMPR=1$, $FMAR=1$, and $AUC=0.5$; (c) Scenario 2 - four genuine minutiae can be correctly extracted, $GMPR=1$, $FMAR=0.3167$, and $AUC=0.8416$; and (d) Scenario 3 - Only one genuine minutiae can be correctly detected, $GMPR=0.25$, $FMAR=0.0667$, and $AUC=0.5916$. The red / yellow hollow points stand for the automatically extracted minutiae and the green solid points represent the genuine minutiae labeled by ground-truth in (a). The ROIs in (c) and (d) are obtained by proposed segmentation module and the method reported in \cite{Cao14} and highlighted by pink convex polygons respectively.}
    \label{show_minutiae_extraction}
\end{figure}
%------------------- show minutiae extraction result -------------------

\begin{table}
\centering
\caption{The comparison of $\overline{GMPR}$, $\overline{FMAR}$ and $\overline{AUC}$ among three scenarios}
\label{GMPR_FMAR_table}
\begin{center}
\begin{tabular}{cccc}
\hline
                    & $\overline{GMPR}$ & $\overline{FMAR}$ & $\overline{AUC}$ \\
\hline
Scenario 1          & 0.9689 & 1 & 0.4845 \\
Scenario 2          & 0.7772 & 0.2711 & 0.7531 \\
Scenario 3          & 0.6854 & 0.2642 & 0.7106 \\
\hline
\end{tabular}
\end{center}
\end{table}

Based on the three scenarios, their corresponding GMPRs and FMARs on $258$ latent images in NIST SD27 are calculated according to the Equation (\ref{gmpr_equ}) and (\ref{fmar_equ}) respectively (shown in Figure \ref{show_minutiae_extraction}). Therefore, the mean GMPR ($\overline{GMPR}$) and the mean FMAR ($\overline{FMAR}$) are summarized in Table \ref{GMPR_FMAR_table}. For the performance comparison among these three scenarios, the area under the curve (AUC) in receiver operating characteristics analysis (ROC analysis) is adopted. Because the GMPR and the FMAR have the equivalent principle as the true positive rate (TPR: sensitivity) and the false positive rate (FPR: specificity) in ROC analysis, the AUC is used to evaluate the impact of ROI-based automated minutiae extraction. That is, the higher AUC, the more reliable the ROI-based automated minutiae extraction performs. For the statistics of the performance regarding to the three scenarios, their corresponding mean AUCs ($\overline{AUC}$) on overall $258$ latent images are calculated. As shown in Table \ref{GMPR_FMAR_table}, the automated minutiae extraction based on the ROI obtained by the proposed segmentation module achieves the best performance, where the genuine minutiae are effectively preserved while the imposter ones are significantly reduced. In addition, the automated minutiae extractor can not detect even one genuine minutiae for some latent images, although the images in such failed cases are completely imported into the automated minutiae extractor. Accordingly, such failed cases where $M{S_1} \cap M{S_2} = \emptyset$ have been evidenced by $\overline{GMPR} = 0.9689 < 1$ in Table \ref{GMPR_FMAR_table}.

\subsection{Experiment 2: Latent Fingerprint Matching}

In this session, the matching experiment is implemented to verify whether the proposed dictionary learning-based ROI segmentation module and the GA-based matching unit are able to boost the matching performance. As the ultimate goal of the fully automated latent fingerprint matching, the matching accuracy plays a vital role to demonstrate the improvement caused by the proposed ROI segmentation module and the GA-based matching unit.

In this experiment, the $258$ latent images provided by NIST SD27 and a background database are used. Such background database are made up of the $258$ mated rolled prints from NIST SD27 and the last $2000$ rolled impressions from NIST SD14. Thus, the size of background database is $2258$. For each query latent print, it is matched against a small-size subset from the established background database. The subset is generated based on the background database and its size is denoted as $R$ ($R<2258$). In such size-$R$ subset, the $R - 1$ rolled prints are randomly and non-repeatedly selected from the background database, while the remaining one is the mate corresponding to the query latent print (the mate can not be included in the pre-selected $R - 1$ prints). Considering the intensive computation involved in the iteration of GA, the subset size $R$ is fixed as $50$ and the parallel computation based on a distributed computers system is utilized. For generalizing to the overall $2258$ rolled prints in the background database, the matching experiment is conducted $10$ times. Besides, the minutiae for the rolled prints and the ROI in latent prints are automatically extracted by VeriFinger SDK. The parameters used in the proposed GA-based matching unit are empirically tuned as follows: the size of chromosomes population $S = 400$, crossover probability $p_c = 0.2$, mutation probability $p_m = 0.05$, minutiae coordinate distance tolerance $\delta_d = 15$, and minutiae orientation difference tolerance $\delta_o = {20^ \circ }$. Further, the value ranges for the affine transformation parameters are set as follows: the orientation range $0 \le \theta  \le {359^ \circ }$, the scaling range $0.8 \le s \le 1.2$, the horizontal shift range $- 400 \le {t_x} \le 400$, and the vertical shift range $- 400 \le {t_y} \le 400$.

\begin{figure}
\centerline{\includegraphics[scale=0.4]{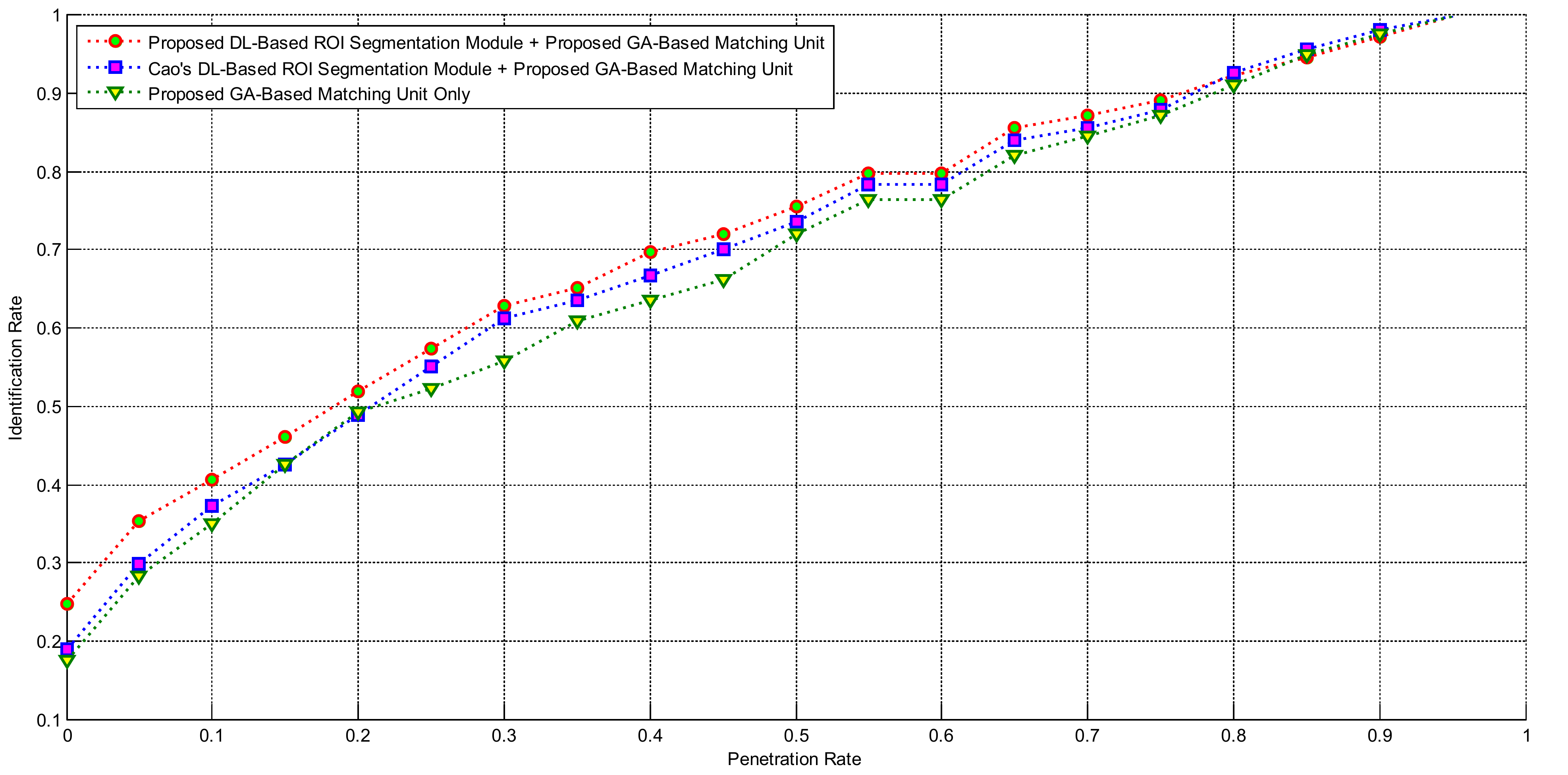}}
\caption{Different $\overline{CMC}$ obtained by the three scenarios.}
\label{CMC_on_Different_Methods}
\end{figure}

In order to evaluate and compare the matching performance, the cumulative match characteristic curve (CMC) is utilized. The CMC indicates the identification rate against the penetration rate $pr$ (e.g. $pr = 1\%, 5\%, 10\%, 15\%, ..., 100\%$). For example, given a query latent print in this experiment, it is searched against $50$ rolled prints. Therefore, the length of candidate list is $50$ and $pr = 10\%$ indicates that the mated rolled print corresponding to the query latent print appears at the $5^{th}$ ($5 = 50 \times 10\%$) of the candidate list. Further, for the statistics of matching performance, the mean CMC ($\overline{CMC}$) and mean penetration rate $\overline{pr}$ on overall $258$ latent prints during $10$ trials are obtained. For the following three scenarios, their corresponding $\overline{CMC}$ and $\overline{pr}$ are demonstrated in Figure \ref{CMC_on_Different_Methods} and Table \ref{mean_pr_table_on_different_methods} respectively.

\begin{itemize}

  \item {\bf Matching Scenario 1 - Proposed GA-Based Matching Unit Only:} without the segmentation module, the minutiae automatically extracted from the whole query latent image are directly imported into the proposed GA-based matching unit.

  \item {\bf Matching Scenario 2 - Proposed ROI Segmentation Module + Proposed GA-Based Matching Unit:} with the preprocessing performed by the proposed ROI segmentation module, the ROI is obtained; then the minutiae automatically extracted from the obtained ROI are imported into the proposed GA-based matching unit.

  \item {\bf Matching Scenario 3 - Cao's ROI Segmentation Method + Proposed GA-Based Matching Unit:} by adopting the approach introduced in \cite{Cao14}, the ROI is obtained; then the minutiae automatically extracted from the obtained ROI are imported into the proposed GA-based matching unit.

\end{itemize}

\begin{table}
\centering
\caption{Different $\overline{pr}$ obtained by the three scenarios}
\label{mean_pr_table_on_different_methods}
\begin{center}
\begin{tabular}{cccc}
\hline
                & Scenario 1       & Scenario 2       & Scenario 3 \\
\hline
$\overline{pr}$ & 38.159\%         & 34.496\%         & 36.434\% \\
\hline
\end{tabular}
\end{center}
\end{table}

As illustrated in Figure \ref{CMC_on_Different_Methods}, the proposed GA-based matching unit is performed as the baseline latent matcher in Scenario 1. Based on such baseline matcher, the further improvement is expected to be achieved via the ROI segmentation method. Scenario 2 and 3 provide the two different segmentation modules for the ROI identification in latent images respectively. As demonstrated in preceding experiment, the reliability of automated minutiae extraction based on the ROI segmentation module in Scenario 2 is better than that of Scenario 3. As a result, the matching performance achieved by Scenario 2 is consequently better than that of Scenario 3. The significant decrease with respect to the mean penetration rate $\overline{pr}$ in Table \ref{mean_pr_table_on_different_methods} also demonstrates the effectiveness of the proposed multi-module latent matcher in Scenario 2.

\begin{figure}
\centerline{\includegraphics[scale=0.4]{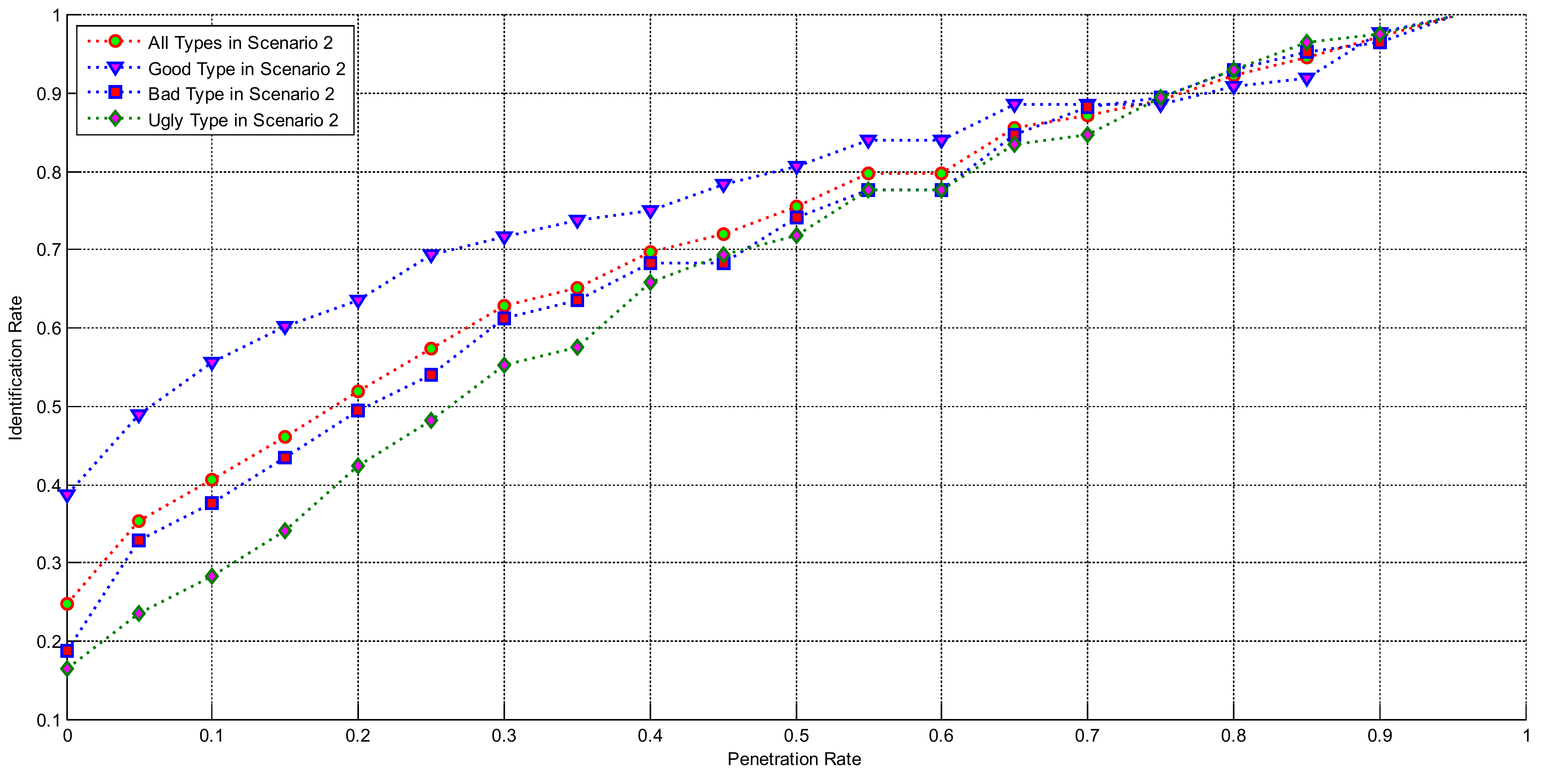}}
\caption{$\overline{CMC}$ obtained in Scenario 2 according to the three categories of latent images: Good, Bad and Ugly.}
\label{CMC_on_Different_Types_by_GA_with_ROI}
\end{figure}

The matching performance on three categories of latent images is further evaluated by Scenario 2. The results shown in Figure \ref{CMC_on_Different_Types_by_GA_with_ROI} and Table \ref{mean_pr_table_on_different_types} demonstrate that $\overline{CMC}$ and $\overline{pr}$ achieved based on the ``Good" latent images is more satisfying rather than the ``Bad" and ``Ugly" ones. The ``Ugly" ones result in the poor matching performance in terms of $\overline{CMC}$ and $\overline{pr}$ because the image quality in ``Ugly" subclass is poor indeed.

\begin{table}
\centering
\caption{$\overline{pr}$ obtained in Scenario 2 according to the three categories of latent images: Good, Bad and Ugly}
\label{mean_pr_table_on_different_types}
\begin{center}
\begin{tabular}{cccc}
\hline
                & Good            & Bad              & Ugly \\
\hline
$\overline{pr}$ & 28.40\%         & 36.17\%         & 39.11\% \\
\hline
\end{tabular}
\end{center}
\end{table}

\section{Conclusions and Future Work}

Although the impressive matching performance has been achieved in the rolled / plain print identification tasks, the matching for the latent fingerprint is still a challenging problem because of the following issues: (i) the poor-quality image and the low-clarity ``ridge-valley" pattern in ROI; and (ii) the corruption with the extra structured image components. Such issues impose the adverse effects on the fully-automated ROI segmentation, minutiae extraction, and minutiae-based matching. In order to deal with the challenges in latent fingerprint matching task, a multi-module latent matching system is introduced in this paper. The proposed latent matcher consists of the following two modules: (i) the dictionary learning-based ROI segmentation scheme; and (ii) the genetic algorithm-based minutiae set matching unit. Experimental results on NIST SD27 latent image database demonstrate the effectiveness of the proposed multi-module latent matching system. In the consideration of its practical value, the proposed matching system can be available in the public domain for latent print identification, and can be also smoothly extended and incorporated with other pre-processing or post-processing modules.

However, the proposed multi-module latent matcher is still not able to handle very low quality latent print images. Since the clarity of the ``ridge-valley" pattern in foreground is poor, the minutiae within such region may not be extracted in a full-automatic mode and consequently the searching against the large-size background database would not be successful. Further, the intensive computation caused by the iteration of GA also needs to be reduced. Therefore, the proposed multi-module latent matching system could be further improved in the following directions:

\begin{itemize}

  \item The development and integration of the enhancement module for the low-clarity ``ridge-valley" pattern in foreground is necessary in the on-going research;

  \item The intensive computation involved in the iteration of GA-based matching unit could be further reduced by defining the simpler fitness function, or exploring to utilize the large-scale parallel or distributed computing system.

\end{itemize}

\end{document}